%% file: paper1.tex
\documentclass[a4paper,fleqn,usenatbib]{mnras}

\usepackage{deluxetable}
\usepackage{newtxtext,newtxmath}
\usepackage[T1]{fontenc}
\usepackage{ae,aecompl}
\usepackage{amssymb, amsmath}
\usepackage{graphicx}
\usepackage{natbib}
\usepackage[usenames, dvipsnames]{color}

\hypersetup{colorlinks=true,
            citecolor=MidnightBlue,
            linkcolor=MidnightBlue,
            filecolor=magenta,      
            urlcolor=cyan}
\DeclareGraphicsExtensions{.pdf,.png,.jpg}

\def\arcsec{{\prime\prime}}

\def\aj{{AJ}}

\def\apj{{ApJ}}
\def\apjs{{ApJS}}
\def\asec{$^{\prime\prime}$}

\def\lax{{$\mathrel{\hbox{\rlap{\hbox{\lower4pt\hbox{${\sim}$}}}\hbox{$<$}}}$}}
\def\gax{{$\mathrel{\hbox{\rlap{\hbox{\lower4pt\hbox{${\sim}$}}}\hbox{$>$}}}$}}
\def\simlt{\lower.5ex\hbox{$\; \buildrel < \over {\sim} \;$}}
\def\simgt{\lower.5ex\hbox{$\; \buildrel > \over {\sim} \;$}}

\def\mnras{{MNRAS}}
\def\nat{{Nature}}
\def\pasp{{PASP}}

\def\sb{mag~arcsec$^{-2}$}
 
\def\msun{$M_\odot$}

\def\etal{{\ et al.~}}

\def\ser{{S\'{e}rsic\ }}
\def\redm{\texttt{redMaPPer}}
\def\cmodel{\texttt{cModel}}
\def\rbcg{\texttt{cenHighMh}}
\def\nbcg{\texttt{cenLowMh}}

\def\mstar{{$M_{\star}$}}
\def\mhalo{{$M_{\mathrm{200b}}$}}
\def\logms{{$\log (M_{\star}/M_{\odot})$}}
\def\logmh{{$\log (M_{\mathrm{200b}}/M_{\odot})$}}

\def\minn{{$M_{\star,10\mathrm{kpc}}$}}
 
\def\mtot{{$M_{\star,100\mathrm{kpc}}$}}

\def\mmax{{$M_{\star,\mathrm{Max}}$}}

\def\mcmodel{{$M_{\star,\mathrm{cModel}}$}}

\def\logmtot{{$\log (M_{\star,100\mathrm{kpc}}/M_{\odot})$}}

\def\logmcmodel{{$\log (M_{\star,\mathrm{cModel}}/M_{\odot})$}}

\def\m2l{{$M_{\star}/L_{\star}$}}
\def\s2n{{$\mathrm{S}/\mathrm{N}$}}
\def\mden{{$\mu_{\star}$}}

\newcommand{\xxx}[1]{\textcolor{red}{\textbf{XXX}}}

\title[Mass Dependent Stellar Halos in Massive Galaxies]{Individual Stellar Halos of 
	   Massive Galaxies Measured to 100 kpc at $0.3<z<0.5$ using Hyper Suprime-Cam}

\author[S. Huang et al.]{
        Song Huang,$^{1,2}$\thanks{E-mail: song.huang@ipmu.jp (SH)}
        Alexie Leauthaud,$^{2,1}$
        Jenny E. Greene,$^{3}$
        Kevin Bundy,$^{4,1}$
        \newauthor
        Yen-Ting Lin,$^{6}$
        Masayuki Tanaka,$^{5}$
        Satoshi Miyazaki,$^{5,7}$
        \newauthor
        Yutaka Komiyama$^{5,7}$
        \\
        $^{1}$Kavli-IPMU, The University of Tokyo Institutes for Advanced Study, 
              the University of Tokyo, Kashiwa 277--8583, Japan\\
        $^{2}$Department of Astronomy and Astrophysics, University of California 
              Santa Cruz, 1156 High St., Santa Cruz, CA 95064, U.S.A\\
        $^{3}$Department of Astrophysical Sciences, Peyton Hall,
              Princeton University, Princeton, NJ 08540, USA \\
        $^{4}$UCO/Lick Observatory, University of California, Santa Cruz,
              1156 High Street, Santa Cruz, CA 95064, USA\\
        $^{5}$National Astronomical Observatory of Japan, 2--21--1 Osawa, Mitaka, 
              Tokyo 181--8588, Japan\\
        $^{6}$Academia Sinica Institute of Astronomy and Astrophysics, 
              P.O. Box 23--141, Taipei 10617, Taiwan\\
        $^{7}$SOKENDAI (The Graduate University for Advanced Studies), Mitaka,
              Tokyo, 181--8588, Japan
        }   
\date{Accepted XXX. Received YYY; in original form ZZZ}        
\pubyear{2017}                                  
  
\begin{document}

\label{firstpage}
\pagerange{\pageref{firstpage}--\pageref{lastpage}}

\maketitle


\begin{abstract} 

    Massive galaxies display extended light profiles that can reach several 
    hundreds of kilo parsecs. 
    These stellar halos provide a fossil record of galaxy assembly histories. 
    Using data that is both wide (${\sim}100$ deg$^2$) and deep ($>28.5$ \sb{} in 
    $i$-band), we present a systematic study of the stellar halos of a sample of more 
    than $3000$ galaxies at $0.3 < z < 0.5$ with $\log M_{\star}/M_{\odot} > 11.4$. 
    Our study is based on high-quality (0.6\asec\ seeing) imaging data from the Hyper 
    Suprime-Cam (HSC) Subaru Strategic Program (SSP), which enables us to individually 
    estimate surface mass density profiles to 100 kpc without stacking. 
    As in previous work, we find that more massive galaxies exhibit more extended outer 
    profiles.  
    When this extended light is not properly accounted for as a result of shallow imaging 
    or inadequate profile modeling, the derived stellar mass function can be 
    significantly underestimated at the highest masses.  
    Across our sample, the ellipticity of outer light profiles increases substantially 
    as we probe larger radii.  
    We show for the first time that these ellipticity gradients steepen dramatically as 
    a function of galaxy mass, but we detect no mass-dependence in outer color gradients. 
    Our results support the two-phase formation scenario for massive galaxies in which 
    outer envelopes are built up at late times from a series of merging events. 
    We provide surface mass surface mass density profiles in a convenient tabulated format 
    to facilitate comparisons with predictions from numerical simulations of 
    galaxy formation.
    
\end{abstract}

\begin{keywords}
    galaxies: elliptical and lenticular, cD --
    galaxies: formation --
    galaxies: photometry -- 
    galaxies: structure -- 
    galaxies: surveys
\end{keywords}


    \begin{figure*}
        \centering 
        \includegraphics[width=\textwidth]{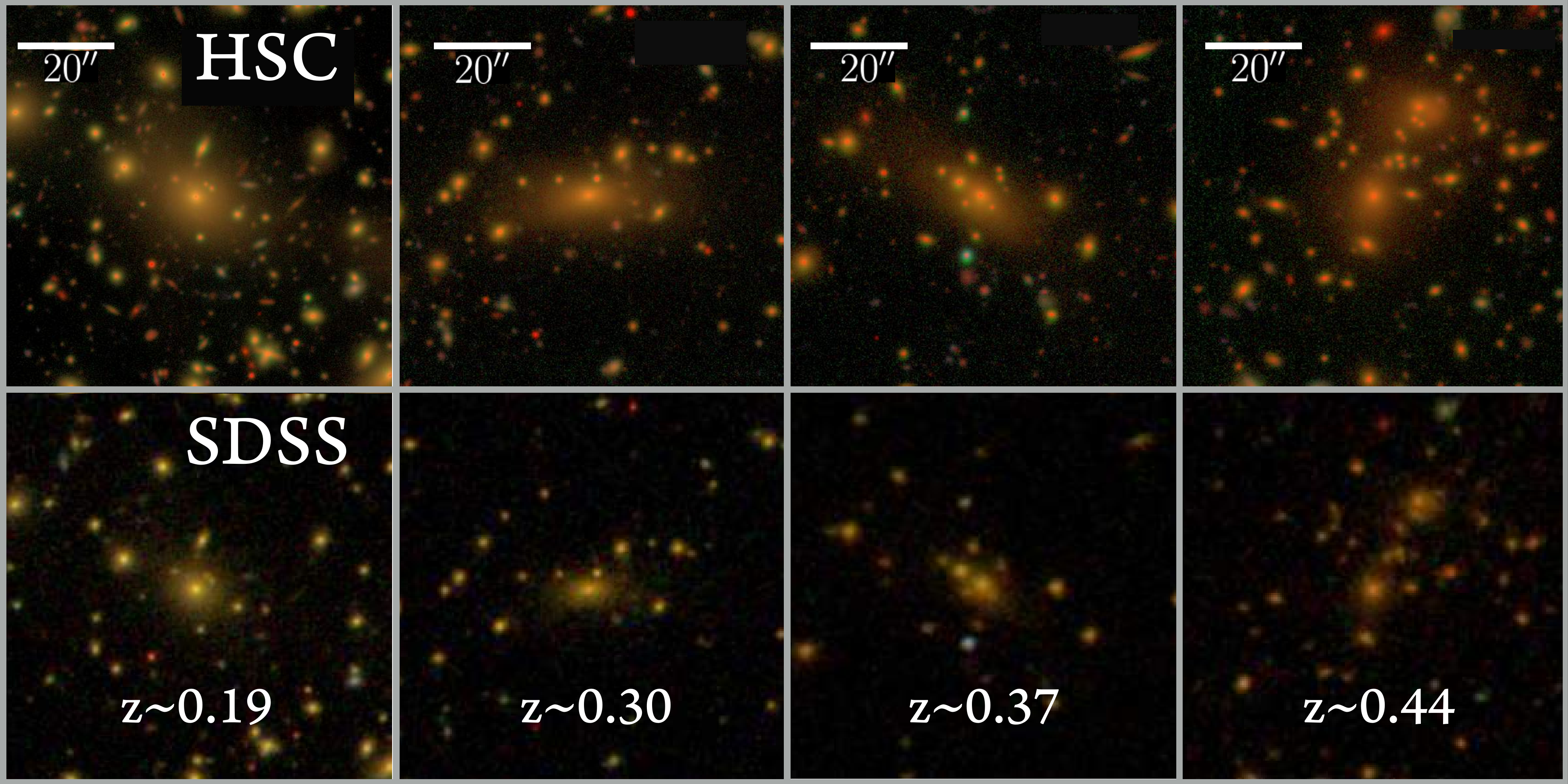}
        \caption{
            A comparison between the depth and imaging quality of SDSS and the HSC wide 
            layer for a sample of nearby massive elliptical galaxies at $0.2 < z < 0.5$.  
            These images are generated using $gri$ band images with an arcsinh stretch 
            \citep{Lupton2004}. 
            The HSC \texttt{WIDE} layer is $3.0$-$4.0$ magnitudes deeper than SDSS.
            This added depth is critical in order map the outskirts of ETGs out to 
            large radii.
            }
        \label{fig:sdss_compare}
    \end{figure*}

\section{Introduction}
    \label{sec:intro}

    Simulations of structure formation within the context of the $\Lambda$-CDM 
    cosmological model make predictions for the hierarchical growth of dark matter 
    halos and galaxies (e.g. \citealt{Baugh1996, DeLucia2006}), but there are many 
    open questions regarding the star-formation history, mass assembly process, and 
    structural evolution of massive galaxies. 
    Massive galaxies are thought to grow according to a ``two-phase'' formation 
    scenario (e.g. \citealt{Oser2010, Oser2012}). 
    In this scenario, the progenitors of $z{\sim} 0$ massive early-type galaxies 
    (ETGs) undergo a rapid growth phase at $z{\sim} 2$ triggered either by disk 
    instabilities or gas-rich mergers (e.g. \citealt{Hopkins2008, Dekel2009}). 
    Observationally, the progenitors of ETGs are thought to correspond to the 
    population of massive quiescent galaxies at high redshift that have smaller 
    average effective radii ($R_{\mathrm{e}}$; e.g. \citealt{Trujillo2006, 
    vanDokkum2008, Cimatti2008}).  
    These high redshift galaxies also have slightly higher central velocity dispersions 
    and stellar mass densities (\mden{}; e.g. \citealt{vandeSande2011, Belli2014}), 
    and more disk-like morphologies (e.g. \citealt{vanderWel2011}) than low redshift 
    galaxies (e.g. \citealt{Bezanson2009, vanDokkum2010}) with similar stellar mass.
    
    Following the initial phase of growth, feedback from stars and/or AGN 
    (e.g. \citealt{Sijacki2007, Fabian2012}) efficiently quenches star formation in 
    massive galaxies. 
    A large fraction of these massive progenitors are already quiescent by $z=1$ 
    (e.g. \citealt{Bezanson2009, Kriek2016}). 
    The second phase of their assembly is driven by non-dissipative processes such 
    as dry mergers with other galaxies (e.g. \citealt{Naab2006, Khochfar2006}). 
    The ``two-phase'' formation of massive galaxies help explain both the 
    observed increase in the effective radii ($R_{\mathrm{e}}$) of massive galaxies 
    (e.g. \citealt{Newman2012, vdWel2014}) and the build-up of stellar 
    halos  (e.g. \citealt{Szomoru2012, Patel2013}). 
    It also posits that minor mergers dominate the mass assembly 
    of massive galaxies at late times 
    (e.g. \citealt{Hilz2012, Hilz2013, Oogi2013, Bedorf2013, Laporte2013}).
    
    Both numerical simulations (e.g. \citealt{Oser2010}) and 
    semi-analytic models (SAM; e.g. \citealt{LeeYi2013, LeeYi2017}) agree that the 
    mass fraction in the accreted component should increase with total galaxy stellar 
    mass (e.g. \citealt{Lackner2012, Cooper2013, Qu2017}).
    For instance, recent results from the 
    Illustris\footnote{\texttt{http://www.illustris-project.org/}} simulation 
    (\citealt{Vogelsberger2014}, \citealt{Genel2014}) predict that the fraction of  
    accreted stars increases significantly and reaches $f_{\mathrm{accreted}}>0.5$ at 
    \logms{}$>11.5$ (\citealt{RodriguezGomez2016})\footnote{The Illustris simulation
    does not reproduce the observed stellar mass function at high \mstar{} end.}. 
    
    Given the success of this two-phase scenario in explaining trends like overall size 
    growth, it is time to confront this model with additional observations, 
    in particular, the detailed surface mass density profiles of low redshift massive 
    galaxies. 
    Early studies based on one-dimensional light profiles found that the surface 
    mass density profiles of nearby ETGs are well described by single-\ser{} profiles 
    (e.g. \citealt{Kormendy2009}; except for the most central region) and that the 
    \ser{} index increases with total luminosity 
    (e.g. \citealt{Graham2013}). 
    However, recently, detailed empirical comparison of surface brightness profiles 
    revealed a more complicated situation showing that ETGs belong to two families, 
    those that follow single-\ser{} law, versus those that significantly deviate from 
    the single-\ser{} profile \citep{Schombert2015}.  
    Two-dimensional analyses have also found that the stellar distributions of massive 
    ETGs are often better fit by multiple-component models 
    (e.g. \citealt{Huang2013a, Oh2017}).  
    \citet{Huang2013b} further suggest a connection between the multi-component 
    nature of massive galaxies and their two-phase assembly history. 
    To further confront the two-phase scenario requires very deep observations of 
    large samples of massive ETGs to correctly estimate their total stellar masses
    (e.g. \citealt{Bernardi2013, DSouza2014}) as well as to quantify the amplitude and 
    scatter among outer envelopes (e.g. \citealt{Capaccioli2015, Iodice2016, 
    Iodice2017}).
    
    Until now, studies of the surface brightness or mass density profiles of massive 
    galaxies have been either conducted using large samples but with shallow imaging, 
    for example the Sloan Digital Sky Survey (SDSS; e.g. \citealt{SDSSDR7, SDSSDR12}), 
    or with deeper imaging but much smaller sample sizes (e.g \citealt{Gonzalez2005}). 
    In this paper, we take advantage of new high-quality (median seeing of
    FWHM${\sim} 0.6^{\arcsec}$ in $i$-band) and deep ($i$-band surface brightness 
    limit $> 28.5$~\sb) images from the Hyper Suprime-Cam (HSC) Subaru Strategic 
    Program \citep[SSP,][]{HSCDR1} to characterize the light profiles of massive 
    galaxies out to 100 kpc. 
    The deep imaging depth and excellent seeing conditions of HSC images make them 
    ideal for mapping the \mstar{} distributions of massive galaxies out to very 
    large radii. 
    We select a large sample (${\sim} 7000$) of massive central galaxies 
    at $0.3 < z < 0.5$ using ${\sim} 100$ deg$^2$ of data from the HSC wide layer. 
    
    
    In this paper, we use this sample to 
    (a) reliably estimate individual surface mass density (\mden{}) profiles of 
    massive galaxies out to 100 kpc, 
    (b) investigate the dependence of their outer stellar halos on total stellar 
    mass, and 
    (c) examine the implications in terms of evaluating the high mass end of the 
    galaxy stellar mass function (SMF). 
    In a second paper in this series, we will investigate the environmental 
    (dark matter halo mass) dependence of the sizes of massive ETGs 
    (Huang\etal in prep.).
    
    This paper is organized as follows. 
    \S~\ref{sec:data} presents our data and initial sample selection. 
    \S~\ref{sec:ellipse} describes our procedure for extracting 1-D surface 
    brightness profiles. 
    \S~\ref{sec:mstar} describes how we estimate stellar mass.
    \S~\ref{sec:final} summaries the final sample selection procedure. 
    Our main results are presented in \S~\ref{sec:result} and discussed in 
    \S~\ref{sec:discussion}. 
    \S~\ref{sec:summary} presents our summary and conclusions.

    Magnitudes use the AB system (\citealt{Oke1983}), and are corrected for Galactic 
    extinction using calibrations from \citet{Schlafly11}.
    In this work, we assume $H_0$ = 70~km~s$^{-1}$ Mpc$^{-1}$, ${\Omega}_{\rm m}=0.3$, 
    and ${\Omega}_{\rm \Lambda}=0.7$.
    Stellar mass is denoted \mstar{} and has been derived using a Chabrier Initial Mass 
    Function (IMF; \citealt{Chabrier2003}).   
      
    Halo mass is defined as $M_{\rm 200b}$ as 
    $M_{\rm 200b}\equiv M(<r_{\rm 200b})=200\bar{\rho} 
    \frac{4}{3}\pi r_{\rm 200b}^3$ where $r_{\rm 200b}$
    is the radius at which the mean interior density is equal to 200 times
    the mean matter density ($\bar{\rho}$). 
    
    Finally, we emphasize that in this work, we do not attempt to disentangle the 
    galaxy light from any ``intra-cluster'' light component (ICL; e.g. 
    \citealt{Carlberg1997, Lin2004, Gonzalez2005, Mihos2005}). 
    Although the rising stellar velocity dispersion in the outskirts of massive 
    brightest cluster galaxy (BCG)
    hints at a kinematically separated ICL component (e.g. 
    \citealt{Dressler1979, Carter1999, Kelson2002, Bender2015, Longobardi2015}),
    it is extremely difficult to reliably isolate them photometrically.
    Moreover, both the stellar halo of main galaxy and the ICL component carry 
    important information regarding the assembly history of central galaxy and its 
    dark matter halo. 
    Therefore, we adopt the view that the light of the main galaxy and the 
    ICL component trace different scales of a single, smooth, and continuous 
    distribution.

    \begin{figure*}
        \centering 
        \includegraphics[width=\textwidth]{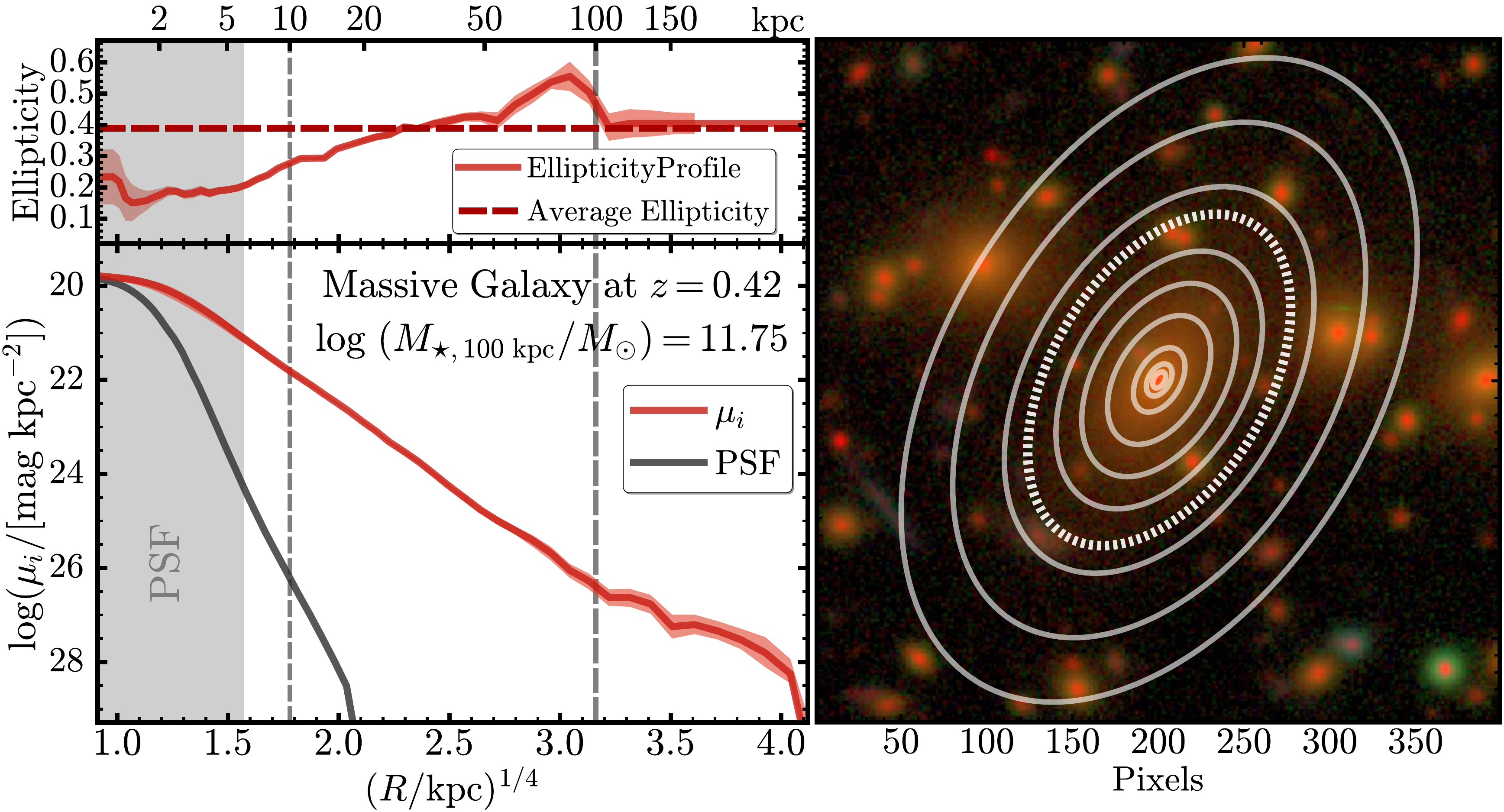}
        \caption{
            Left: example of the 1-D surface brightness and ellipticity profile 
            of a massive galaxy at $z=0.23$ in the $i$-band extracted using 
            \texttt{Ellipse}. 
            In this work, we always show the radial profile using a $R^{1/4}$ scaling 
            on the x-axis.
            By using this scale, the de Vaucouleurs profile will appear as a straight 
            line on this figure.  
            We also plot the relative brightness profile of the PSF model normalized 
            at the central surface brightness of the galaxy to highlight the region 
            most strongly affected by seeing.
            The grey shaded region highlights the region ($r<6$ kpc) that is equivalent 
            to twice the size of the half-width of a 1\asec{} seeing at $z\sim 0.5$.
            It is a very conservative estimate of the region that we can not reliable 
            extract 1-D profile due to the smearing effect of seeing.
            On the top panel, the dashed line shows the mean ellipticity used for the 
            final isophote. 
            Right: the three color image of this galaxy with isophotes 
            extracted by \texttt{Ellipse}. 
            The thick dotted line highlights the isophote with 
            $\mu_{i}{\sim} 28.5$~\sb.
            }
        \label{fig:ellipse}
    \end{figure*}

\section{Data And Sample Selection}
    \label{sec:data}

\subsection{The Hyper Suprime-Cam Survey}
    \label{ssec:hsc}

    The Subaru Strategic Program (SSP, \citealt{HSC_SSP, HSC_DR1}) makes use of the 
    new prime-focus camera, the Hyper Suprime-Cam (HSC;~\citealt{Miyazaki2012}, 
    Miyazaki in~prep.), on the 
    8.2-m Subaru telescope at Mauna Kea. 
    The ambitious multi-layer HSC survey takes advantage of the large field of 
    view (FoV;~1.5 deg in diameter) of this camera and will cover $>1000$ deg$^2$ 
    of sky in 5 broad bands ($grizy$) to a limiting depth of $r {\sim} 26$ mag 
    in the \texttt{WIDE} layer. 
    This work is based on the internal data release \texttt{S15B}, which covers 
    ${\sim} 110$ deg$^2$ in all 5-band to full \texttt{WIDE} depth.  
    The regions covered by this release overlap with a number of spectroscopic surveys 
    (e.g.\ SDSS/BOSS: \citealt{Eisenstein2011}, \citealt{Alam2015}; 
    GAMA: \citealt{Driver2011}, \citealt{Liske2015}).
    \texttt{S15B} release has similar sky coverage with the Public Data Release 1
    (Please see Table 3 in \citealt{HSC_DR1} for detailed comparison).

    The HSC \texttt{WIDE} survey is about $3.0$-$4.0$ magnitudes deeper in terms of 
    the $i$-band surface brightness limit than SDSS. 
    Combined with the excellent imaging resolution (the median $i$-band seeing is 
    0.6\asec) and the wide area, the HSC survey represents an ideal dataset to perform 
    statistical studies of the surface brightness profiles of massive galaxies out to 
    their distant outskirts.  
    Fig~\ref{fig:sdss_compare} illustrates the quality of HSC imaging compared to SDSS 
    for three low redshift ETGs, and shows that HSC survey data are well suited for 
    mapping the stellar distribution of massive galaxies out to large radii.

	HSC $i$-band images typically have the best seeing compared to other bands because 
	of strict requirements driven by weak lensing science. 
    We will therefore use $i$-band images to measure the stellar distributions of 
    massive galaxies.
    
\subsection{HSC Data Processing}
    \label{sec:pipeline}

    The full details of the HSC data processing can be found in \citet{Bosch2017}
    and are briefly summarized here. 
    The HSC SSP data are processed with \texttt{hscPipe 4.0.2}, a derivative of the 
    Large Synoptic Survey Telescope (LSST) pipeline (e.g.\ \citealt{Juric2015}; 
    \citealt{Axelrod2010}), modified for HSC. 
    \texttt{hscPipe} first performs a number of tasks at the single exposure level 
    (bias subtraction, flat fielding, background modeling, object detection and 
    measurements). 
    Astrometric and photometric calibrations are performed at the single exposure 
    level. 
    \texttt{hscPipe} then warps different exposures on to a common World Coordinate 
    System (WCS) and combines them into coadded images. 
    At this stage, \texttt{hscPipe} updates the images with a better astrometric and 
    photometric calibration using stars that are common among exposures. 
    
    The pixel scale of the combined images is $0.168$\asec{}. 
    Photometric calibration is based on data from the Panoramic Survey Telescope 
    and Rapid Response System (Pan-STARRS) 1 imaging survey 
    (\citealt{Schlafly2012}, \citealt{Tonry2012}, \citealt{Magnier2013}). 
    To achieve consistent deblending and photometry across all bands, \texttt{hscPipe} 
    performs multi-band post-processing at the \texttt{coadd} level. 
    First, \texttt{hscPipe} performs object detection on \texttt{coadd} images in 
    each band independently and records the flux peak and the above-threshold region 
    (referred as a \texttt{footprint}) for each source. 
    Next, \texttt{footprints} and peaks from different bands are merged together before     
    performing deblending and measurements. 
    Finally, \texttt{hscPipe} selects a reference band for each object based on the 
    $S/N$ in different bands (for most galaxies in this work, the reference band is 
    the $i$-band). 
    After fixing the centroids, shape, and other non-amplitude parameters of each 
    object in this reference catalog, \texttt{hscPipe} performs forced photometry 
    on the \texttt{coadd} image in each band. 
    This forced photometry approach is optimized to yield accurate galaxy colors
    at $i_{\mathrm{CModel}\leq25.0}$ mag (see \citealt{SynPipe}).
       
    For each galaxy, \texttt{hscPipe} measures a \cmodel{} magnitude using an approach 
    that is similar to SDSS (\citealt{Bosch2017}). 
    However, as opposed to SDSS, the HSC \cmodel{} is based on forced multi-band 
    photometry which means that it can accurately measure both the 
    \textit{fluxes and colors of galaxies}. 
    The HSC \cmodel{} algorithm fits the flux distribution of each object using a 
    combination of a de~Vaucouleur and an exponential component and accounts for the PSF. 
    The performance of this algorithm has been tested using synthetic objects 
    (\citealt{SynPipe}), and the results indicate that, generally speaking, 
    the HSC \cmodel{} photometry is accurate down to $i >25.0$ mag.  
    However, \cmodel{} currently systematically underestimates the total fluxes of 
    massive ETGs with extended stellar distributions. 
    This is caused by an intrinsic limitation of \cmodel{} as it is incapable of
    modeling profiles with extremely extended outskirts, a problem that is exacerbated 
    at the depth of the HSC survey. 
    In addition, at the depth of the HSC survey, accurate deblending in the vicinity of
    large ETGs where satellites and background galaxies often blend with the low surface 
    brightness stellar envelope is a challenging problem. 
    The deblending method currently implemented in \texttt{hscPipe} tends to 
    ``over-deblend'' the outskirts of bright galaxies and leads to an 
    under-estimation of the total flux of massive ETGs (this is discussed further in 
    \citealt{Bosch2017}).  
    For these reasons, our results will be based on custom-developed code to measure 
    the luminosities and stellar masses of massive galaxies. 
    We use the HSC \texttt{hscPipe} photometry for two purposes: 
    1) to perform a first broad sample selection, and 2) to estimate the average 
    color of massive galaxies.

  \begin{figure*}
      \centering 
      \includegraphics[width=\textwidth]{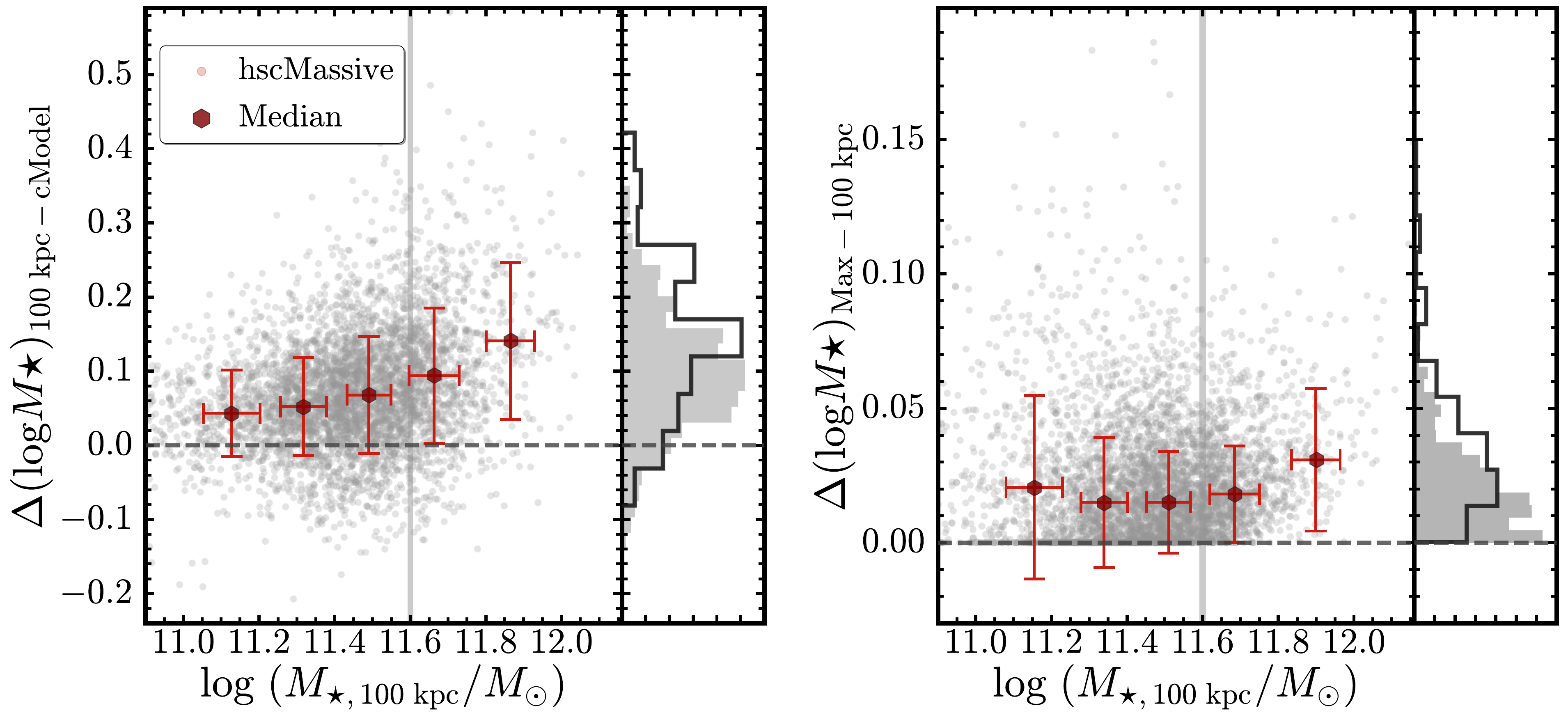}
      \caption{              
          \textbf{Left:} Difference between \mcmodel{} and \mtot{} for massive
      	  galaxies (grey dots).
      	  The running-median of the mass difference is shown by large red hexagons. 
      	  On average, \mcmodel{} underestimates the total stellar mass of massive 
          galaxies by 0.1 dex while in some cases, the difference can exceed 0.2 dex.
          Vertical histograms indicate the mass difference for all galaxies (shaded 
          histogram) and for the ones with \logmtot{}$>11.6$ (empty histogram).
          \textbf{Right:} Difference between \mmax{} and \mtot{} in the same format. 
          The average difference is small (0.02 dex) and with no clear mass--dependence. 
          \textbf{Please note that the scales of the vertical axes are different 
            for these two figures.}
          }
      \label{fig:mass_diff}
  \end{figure*}
          
    
\subsection{Initial Massive Galaxy Sample}
    \label{ssec:initial}
    
    We begin by using a broad flux cut to select an initial sample of massive 
    galaxies at $z < 0.5$ from the HSC photometric catalog. 
    Based on \citet{Leauthaud2016}, $i_{\mathrm{SDSS, cModel}} \leq 21.0$ mag can 
    define a sample that includes almost all \logms{}$\geq 11.5$ galaxies.    
    We therefore perform an initial conservative selection of massive galaxies
    with $i_{\mathrm{HSC, cModel}} \leq 21.5$\footnote{We neglect small differences
    between the response curves of the SDSS-$i$ and HSC-$i$ filters.}. 
    We also limit our sample to regions that have reached the required depth of 
    the \texttt{WIDE} survey in $i$-band as defined in \citet{HSC_DR1}.
    
    We further select extended objects with no deblending errors, with well defined 
    centroids, and with useful \texttt{cModel} magnitudes in all five bands. 
    After removing objects that have pixels affected by saturation, cosmic-rays, or 
    other optical artifacts\footnote{each criterion removes less than 8\% of the 
    entire sample.}, this sample corresponds to 1760845 galaxies and will be referred 
    as \texttt{hscPho}. 
        
    Here we limit our study to the very high-mass end where the majority of galaxies 
    have either a spectroscopic redshift or a robust red-sequence photo-$z$ from the 
    \redm{} galaxy cluster catalog\footnote{See: http://risa.stanford.edu/redmapper/} 
    (e.g.\ \citealt{Rykoff2014}; \citealt{Rozo2015b}).  

    We match the \texttt{hscPho} sample with a spec-$z$ catalog compiled by the HSC 
    team.
    It is created by matching HSC objects with a series of publicly available 
    spectroscopic redshifts (e.g.\ SDSS DR12 \citealt{SDSSDR12}; 
    GAMA DR2 \citealt{Liske2015}). 
    The spec-$z$ quality flags from different catalogs are homogenized into a single 
    flag that indicates secure redshifts.
    Please see \S~4.4.2 of \citet{HSC_DR1} for details of this catalog.  
    To ensure reasonable \mstar{}-completeness at the high-\mstar{} end we focus on
    the redshift range $0.3 \leq z \leq 0.5$. 
   
    Objects without a spectroscopic redshift are matched with central 
    galaxies from the \redm{} SDSS DR8 (\citealt{Rykoff2014}) catalog using a 
    $2.0^{\arcsec}$ matching radius. 
    Matched objects with a red-sequence photo-$z$ ($0.3 \leq z_{\lambda} < 0.5$) are 
    included in our sample. 
    The accuracy of the red-sequence photo-$z$ is sufficient (median 
    $|z_{\lambda} - z_{\mathrm{Spec}}| {\sim} 0.01$) for our purpose.
    The \redm{} catalog provides an additional 133 unique redshifts for massive 
    galaxies in our sample.
        
    In total, at $0.3 \leq z \leq 0.5$, our sample consists of 25286 galaxies with 
    reliable redshift information (referred as \texttt{hscZ}).
    The majority of our redshifts comes from the BOSS and SDSS ``legacy'' LRG samples. 
    The GAMA survey provides an additional $14$\% of all spectroscopic redshifts.
    Although the GAMA survey only covers parts of the \texttt{S15B} data release, 
    hence affects the homogeneity of our sample, it will not affect the results of 
    this work.
    We will discuss this more in \S \ref{sec:final}.

    We choose the redshift range $0.3 \leq z \leq 0.5$ to make sure that: 
    (1) The inner region of massive galaxies can be resolved, and \mstar{} within 
    10 kpc can be reliably measured; 
    (2) The background noise and cosmological dimming are not major issues so that the 
    \mden{} profile can be measured out to $>100$ kpc; 
    (3) Redshift evolution in the stellar population properties can be largely 
    ignored.  
    Also at higher redshift, the completeness of the spec-$z$ sample starts to decline. 
    And finally, the over-subtraction of the background level becomes a more serious 
    issue at lower redshifts.  
   
    We will now describe our one-dimensional photometric analysis 
    (\S \ref{sec:ellipse}) and our stellar mass estimates (\S \ref{sec:mstar}). 
    We will then define the final sample in \S \ref{sec:final}.

\section{Measurements of 1-D Surface Brightness Profiles}
    \label{sec:ellipse}
    
    The surface brightness profiles of massive ETG are not well modeled by the 
    de~Vaucouleurs or single-\ser law, especially at the imaging depth of HSC.
    These models will fail to simultaneously describe the profile in both the inner 
    and the outer regions and also cannot account for any radial variations in 
    ellipticity and position angle. 
    In principle, massive galaxies can still be described by more complex 
    models (e.g \citealt{Huang2013a, Huang2013b, Oh2017}), but the results are still 
    sensitive to the choice of a particular model (e.g. de~Vaucouleurs or \ser{} 
    profile), the number of components, initial guesses of parameters, and internal 
    degeneracies among different parameters. 
    Background subtraction uncertainties can also affect the 2-D model fitting method, 
    especially for the massive ETGs that make up our sample. 
    
    We therefore perform elliptical isophote fitting using the IRAF \texttt{Ellipse} 
    algorithm (Jedrzejewski 1987) to estimate the total luminosities of massive 
    galaxies and to measure their one-dimensional stellar mass surface 
    density profiles (\mden{}). 
    This 1-D method is less affected by the issues mentioned above. 
    Also, we will only study galaxies in the radial range where we are less 
    sensitive to either the PSF or the background subtraction.
    We ignore the inner $\sim6$ kpc, which is twice the size of 1\asec{} seeing at 
    $z=0.5$.
    Using this conservative choice, we can safely ignore the smearing effect of 
    seeing outside this radius.
    As we will show later, we confirm this by comparing our HSC profiles with 
    observations with higher spatial resolution. 
    As for the impact from background subtraction, we will focus on the profiles 
    within 100 kpc. 
    This is an empirical, but also conservative choice based on the tests we 
    conducted on background-corrected postage-stamps. 
    Once the surrounding objects are appropriately masked out, the extracted 1-D 
    surface brightness profiles rarely see unphysical truncation or fluctuation 
    within 100 kpc, especially for the \logmtot{}$>11.6$ galaxies. 
    Please see Appendix~\ref{app:ellipse} for more details on these tests.
        
    We prepare large $i$-band postage-stamps for each galaxy that extend to 750 kpc 
    in radius, along with a bad pixel mask and the PSF model. 
    These postage-stamps include all of the light of the galaxy and are also large 
    enough to evaluate the background level. We choose to use $i$-band images 
    because they trace the stellar mass distributions of massive galaxies at 
    $0.3 \leq z \leq 0.5$ reasonably well 
    (the observed $i$-band corresponds to a rest-frame $g$ or $r$ band), but also 
    because they have better seeing and much lower background levels than the $z$ 
    and $y$ band images (although these in principle may be better tracers of 
    \mden{}). 
    
    For each cut-out, to overcome the \texttt{hscPipe} ``over-deblending'' issue, 
    we use a customized procedure to detect and aggressively mask out
    neighbouring objects. 
    Furthermore, \texttt{hscPipe} tends to over-subtract the background around 
    bright objects. 
    To improve the background subtraction, we first aggressively mask 
    out all objects (including the central massive galaxy), and derive an 
    empirical background correction using \texttt{SExtractor}.
    These procedures are described in detail in Appendix~\ref{app:ellipse}. 
    We should point out that we are not using the photometric results from our 
    customized process, but simply rely on them for improved local background model
    and appropriate object mask.

    Then, we run \texttt{Ellipse} on the background-corrected, masked cut-outs 
    following the methodology of \citet{Li2012}. 
    In short, we first fit each isophote using a free centroid and shape 
    (ellipticity and position angle). 
    We then fix the centroid (using the mean flux-weighted centroid) and estimate
    the mean ellipticity and position angles of all isophotes. 
    Finally, we extract a 1-D surface brightness profile along the major axis using 
    the mean ellipticity and position angle. 
    We correct these surface brightness profiles for Galactic extinction and 
    cosmological dimming, and integrate them to various radii to get the luminosity 
    within different physical (elliptical) apertures. 
    Fig~\ref{fig:ellipse} shows an example of the 1-D surface brightness and 
    ellipticity profile for a massive galaxy at $z{\sim}0.2$ and also highlights 
    a few isophotes.    

    We test our procedure using different mask sizes, different \texttt{Ellipse} 
    parameters, and with or without our background correction. 
    Based on these tests, we find that our 1-D surface brightness profiles are reliable 
    up to surfaces brightness levels of $i{\sim}28.5$ \sb. 
    Beyond that, some of our profiles shows signs of truncation and/or large 
    fluctuations which are due to either the uncertainty in the background 
    subtraction or the unmasked flux from other objects.
    We choose to limit our study to surface brightness levels up to ${\sim} 28.5$ \sb. 
    This is a conservative choice but already enables us to measure light profiles 
    out to $100$ kpc on a galaxy-by-galaxy basis (no stacking). 
    For more technical details of the \texttt{Ellipse} procedure, please see 
    Appendix~\ref{app:ellipse}.

    
    We cannot extract reliable 1-D profiles for a small fraction of massive galaxies 
    because they are heavily masked out for either physical 
    (e.g.\ late-stage major merger) or nuisance (e.g.\ nearby foreground 
    galaxy or bright star) reasons. 
    This is an intrinsic limitation of the 1-D method, and it removes ${\sim}10$\% of 
    the sample.
    We visually examine the 3-color images of randomly selected galaxies with failed 
    1-D profiles.  
    Most of them are relative small galaxies that are severely contaminated by nearby 
    objects, and will not affect the results of this work.  
    Meanwhile, it is worth noting that this does exclude most major merging systems 
    among massive galaxies. 
        
    Given exquisite profiles extending to 100 kpc, the definition and meaning 
    of ``total'' magnitude (and stellar mass) becomes nuanced.
    With the help of the average \m2l{} estimated in the next section 
    (\S \ref{sec:mstar}), we integrate the profile to a range of radii, and estimate 
    the stellar mass within these different projected 2-D apertures.  
    Motivated by the two-phase scenario, we will consider two benchmark physical 
    apertures throughout this work:
    
    \begin{itemize} 
       
        \item \textbf{The \mstar{} within the inner 10 kpc} 
            (hereafter noted \minn{}). 
            According to the two-phase scenario, the \textit{in situ} star-formation 
            phase quickly builds up the inner, dense core of massive ETGs.  
            Based on recent observation (e.g.~\citealt{vanDokkum2010}) and 
            simulations (e.g. \citealt{RodriguezGomez2016}), the \textit{in situ} 
            component dominates the \mstar{} within one effective radius 
            ($R_{\mathrm{e}}$, or 5-10 kpc) of $z{\sim}0$ massive ETGs.
            We therefore use \minn{} as a proxy for the mass formed during the 
            \textit{in situ} phase. 
            Given the quality of the HSC data, we can reliably measure \minn{} over 
            our redshift range ($1.0^{\arcsec}$ in radius equals 4.4 and 6.1 kpc 
            at redshifts 0.3 and 0.5 respectively).  
            It is worth noting that, at the very high-\mstar{} end, accreted stars 
            may make a significant contribution to the mass within 10 kpc 
            (e.g. \citealt{RodriguezGomez2016}). 
            We will further discuss the justification of this assumption in 
            \ref{ssec:twophase}.
            
        \item \textbf{The stellar mass within 100 kpc} 
            (hereafter noted \mtot{}). 
            For our galaxy sample, a 100 kpc aperture corresponds to 5-10 
            $\times R_{\mathrm{e}}$. 
            We show in \S~\ref{ssec:mtotal} that most of the total \mstar{} for 
            these ETGs lies within a 100 kpc radius and that \mtot{} is 
            a good proxy for the ``total'' \mstar{}. 
            Although not perfect, we argue that our measurements of \mtot{} (which are 
            actually measuring the light directly out to 100 kpc) are a better tracer 
            of total \mstar{} than model-dependent results from shallower data 
            (such as SDSS) which rely on extrapolating the light profiles of galaxies 
            out to large radii.
            
       \end{itemize}
       
    We should point out that both \minn{} and \mtot{} are measured after adopting a 
    isophote with fixed ellipticity and position angle. 
    The 10 kpc and 100 kpc here refer to the radius along the major axis of the 
    elliptical isophote. 
       
    
\section{Stellar Masses and Mass Density Profiles}
    \label{sec:mstar}
    
\subsection{Stellar Masses from SED Fitting}
    \label{ssec:isedfit}
   
    To convert luminosities into \mstar{}, we assume that these massive galaxies 
    can be well described by an average \m2l{}. 
    This is a reasonable assumption considering that they are mostly dominated by 
    old stellar populations and are known to have only shallow color gradients. 
    We will further justify this point by measuring their median color profiles in 
    \S~\ref{ssec:ell_color}.

    We use the broadband Spectral Energy Distributions (SEDs) fitting 
    (see \citealt{Walcher2011} for a recent review) code 
    \texttt{iSEDFit}\footnote{http://www.sos.siena.edu/~jmoustakas/isedfit/} 
    (\citealt{Moustakas13}) to estimate the average \m2l{} and $k$-corrections using
    5-band HSC \cmodel{} fluxes.
    Although \cmodel{} tends to underestimate the total fluxes of bright, extended 
    objects, it can still yield accurate \emph{average} colors thanks to the 
    forced-photometry method that takes the PSF convolution into account
    (e.g. \citealt{SynPipe}). 

    \texttt{iSEDFit} takes a simplified Bayesian approach. 
    In short, it first generates a large grid of SEDs from synthetic stellar 
    population models by drawing randomly from the prior distributions of relevant
    parameters (e.g. age, metallicity, dust extinction, and star formation history).
    Based on these models, it uses the observed photometry and redshift to compute 
    the statistical likelihood, and generates the posterior probability distribution 
    functions (PDF) for each parameter.  
    To get the best estimate of a given parameter, \texttt{iSEDFit} integrates the 
    full PDF over all the other nuisance parameters.
    Then, the median value and the 1-$\sigma$ uncertainty are derived based on the 
    marginalized PDF. 
    Please refer to \citet{Moustakas13} for technical details. 
    
    In this work, we derive average \m2l{} using the Flexible Stellar Population 
    Synthesis\footnote{http://scholar.harvard.edu/cconroy/sps-models}
    (FSPS; \texttt{v2.4}; \citealt{FSPS}, \citealt{Conroy2010}) model based on the 
    MILES\footnote{http://www.iac.es/proyecto/miles/pages/stellar-libraries}
    (\citealt{MILES1}, \citealt{MILES2}) stellar library and assuming a
    \citet{Chabrier2003} IMF between 0.1 to 100 \msun. 
    The star formation history (SFH) is assumed to follow a delayed-$\tau$ model with 
    stochastic star bursts (see Appendix~\ref{app:sed}). 
    This SFH is appropriate for massive galaxies at low redshifts 
    (e.g.\ \citealt{Kauffmann2003}). 
    For stellar metallicity 
    ($[\mathrm{M}/\mathrm{H}]=\log\ (\mathrm{Z}/\mathrm{Z}_{\odot})$), we assume a 
    flat distribution between 0.004 to 0.03 (the highest value allowed by FSPS).
    We adopt the \citet{Calzetti2000} extinction law with a order two Gamma 
    distribution of $A_{V}$ between 0 to 2 magnitudes. 
    The majority of our galaxies are red and quiescent so the results are not 
    very sensitive to parameters related to the SFH or the internal dust extinction. 
    To achieve reasonable sampling across these parameters, we generate 250000 
    models. 
    
    We construct five-band SEDs using the forced-photometry \cmodel{} magnitudes 
    corrected for Galactic extinction. 
    Presently, \cmodel{} only accounts for the statistical error on the flux 
    measurement and it certainly underestimates the true flux errors of bright 
    galaxies.  
    For this work, we supply \texttt{iSEDFit} with simplified flux errors assuming 
    $S/N = 100$ for the $riz$ bands, and $S/N = 80$ for the $g$ and $y$ band 
    (on average, images in $gy$ bands are shallower in depth and/or have higher 
    background noise).  
    These empirical \s2n{} choices still only provide lower-limits of the true 
    systematic uncertainties from the model-fitting process. 
    In \citealt{SynPipe}, we evaluate the accuracy of HSC \cmodel{} photometry 
    using synthetic galaxies, and show that \cmodel{} provides excellent measurements 
    of five-band colors, which are crucial for reliable \m2l{} estimates.
    The typical uncertainty of \logms{} is around 0.06-0.08 dex at \logms${\sim} 11.5$.
    
    In Appendix~\ref{app:sed}, we briefly summarize the basic statistics of 
    the sample by showing the relationships between \mtot{} and stellar age, 
    metallicity, and internal dust extinction. 
    All these properties behave reasonably for massive galaxies in this sample. 
    Using the $k$-corrected optical color, we can also confirm that the sample follows 
    a tight ``red-sequence''.  
    Please see Appendix~\ref{app:sed} for further details.
      

  \begin{figure}
      \centering 
      \includegraphics[width=\columnwidth]{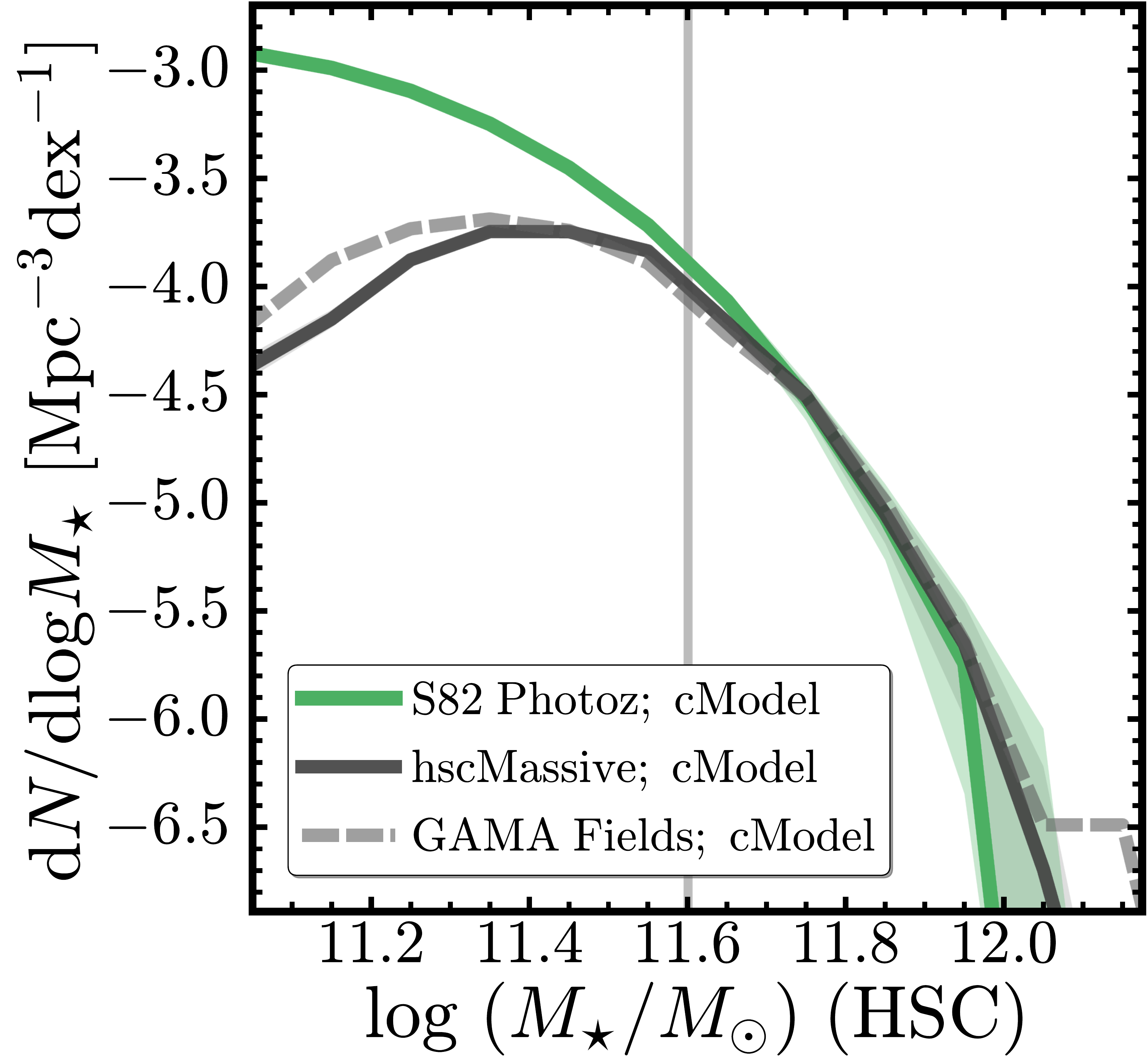}
      \caption{
          Evaluation of the \mstar{} completeness of the HSC massive galaxy sample.
          We compare the volume number density function of the massive galaxies 
          for this work (black line) with the one of a much more complete sample
          from the S82-MGC catalog (green line). 
          The grey dashed line shows the number density function of HSC massive 
          galaxies in the three GAMA fields for comparison.
          The associated uncertainties derived from bootstrap resampling are shown in 
          shaded regions. 
          The vertical grey line highlight the \logms{}$=11.6$ limit.  
          Below it, the HSC massive galaxy sample becomes significantly incomplete in 
          stellar mass. 
          }
      \label{fig:mass_complete}
  \end{figure}     

\subsection{``Total'' Stellar Masses}
    \label{ssec:mtotal}
    
    Using the best-fit \mstar{} from \texttt{iSEDFit} (noted as as \mcmodel{}), 
    we estimate the average \m2l{} in the $i$-band, then use that \m2l{} to convert 
    our 1-D luminosity density profiles into stellar mass density (\mden{}) profiles. 
    We also convert our 10 and 100 kpc aperture luminosities into corresponding stellar 
    mass estimates (noted as \minn{} and \mtot{}).

    For the remainder of this paper, we will use \mtot{} as a proxy of ``total'' 
    stellar mass. 
    As expected, the integration of the 1-D profile out to very large radius recovers 
    more luminosity (stellar mass) compared to the \cmodel{}-based estimates 
    (Fig \ref{fig:mass_diff}).
    At the high-\mstar{} end (e.g. \logmtot{}$>11.6$), the average difference is larger 
    than 0.1 dex and can be as large as 0.2--0.3 dex.  
    More importantly, the differences between \mtot{} and \mcmodel{} clearly show a 
    dependence on total stellar mass as \mcmodel{} tends to miss more light in 
    more massive galaxies.  
    This relates to the mass-dependent nature of the stellar halos of massive 
    galaxies and the intrinsic limitation of \cmodel{} method, which we will discuss
    more in the next section.
    Such differences also have important implications for estimates of the stellar 
    mass function and for studies of the environment-dependence of galaxy structure. 
    These topics will be discussed in \S~\ref{sec:discussion}.
    
    Although 100 kpc is already a very large radius that should enclose the majority 
    of stars that belong to these massive galaxies, we know that their \mden{} 
    profiles extend beyond 100 kpc without showing any signs of truncation
    (e.g. \citealt{Gonzalez2005, Tal2011, DSouza2014}).
    Therefore, for massive galaxies, even \mtot{} should be only considered a lower
    limit on the ``total'' \mstar{}. 
    In Fig \ref{fig:mass_diff}, we integrate the \mden{} of each galaxy to the 
    edge of the postage-stamp, and pick the isophote that gives us the highest  
    luminosity to estimate the \mstar{} within and call this one \mmax{}. 
    The right panel of Fig \ref{fig:mass_diff} compares \mmax{} and \mtot{}.  
    Uncertainties in the background subtraction, and the impact of neighbouring 
    objects, means that \mmax{} is much more uncertain than \mtot{}. 
    Nonetheless, we do see that \mmax{} is larger than than \mtot{}. 
    However, the mass differences are on average very small ($\sim0.02-0.03$ dex) and 
    do not show a strong mass dependence.  
    This confirms that, at the current depth of HSC images, \mtot{} can be used as 
    a good proxy of ``total'' stellar mass.   
    
    Our methodology ignores radial variations in \m2l{}. 
    It is well known that massive ETGs have negative optical color gradients 
    indicating gradients in \m2l{} (e.g.\ \citealt{Carollo1993, Davies1993, 
    LaBarbera2012, DSouza2015}). 
    Assuming all massive galaxies in our sample have negative color gradients, and 
    there is a simple monotonic relation between optical color and \m2l{}, the
    average \m2l{} we used should in principle underestimate the \mstar{} in the 
    center while overestimate the \mstar{} in the outskirt.
    However, these color gradients are shallow and smooth out to a few times the 
    effective radius (e.g. \citealt{LaBarbera2010, Tal2011, DSouza2014}, 
    color gradients at larger radii are not yet well quantified). 
    Because the gradients are shallow, using an average \m2l{} is unlikely to bias 
    our results on \mstar{} measurements.  
    In \citet{Huang2016}, the authors conduct multi-band decomposition for a sample 
    of very nearby elliptical galaxies, and estimate the \m2l{} of each component 
    separately. 
    The sum of all components suggest a slightly higher \mstar{} ($0.05-0.10$ dex 
    when typical uncertainty of \mstar{} is 0.12-0.15 dex) and the mass differences 
    show no dependence on \mstar{}. 
 
    Color gradients will be discussed more in \S~\ref{ssec:ell_color}. 
    In summary, our results about the mass dependence of \mden{} profiles should not 
    be affected by the assumption of a constant \m2l{} ratio because optical color 
    gradients in our sample do not show a dependence on stellar mass.
    
    
    
        
   

\subsection{Stellar Mass Completeness}
    \label{ssec:complete}
    
    With the help of the Stripe82 Massive Galaxy Catalog (\texttt{S82-MGC},
    \citealt{Bundy2015}
    )\footnote{http://www.ucolick.org/\~{}kbundy/massivegalaxies/s82-mgc.html}, 
    we investigate the \mstar{} completeness of our samples. 
    The \texttt{S82-MGC} sample matches the deeper SDSS photometric data in the 
    Stripe 82 region (\citealt{Annis2014}) with the near infrared data from the United 
    Kingdom Infrared Telescope Infrared Deep Sky Survey (UKIDSS; 
    \citealt{Lawrence2007}). 
    Comparing to normal SDSS images, the deeper photometry and better photo-$z$s 
    from \texttt{S82-MGC} make this sample complete to \logms{}$\geq 11.2$ at $z<0.7$ 
    which makes it sufficient to evaluate the completeness of our HSC sample.   
	By comparing with the \texttt{S82-MGC} sample (\citealt{Bundy2015}), 
	\citet{Leauthaud2016} have measured the \mstar{} completeness of the combined BOSS 
	and SDSS samples. 
    They estimate that the BOSS spec-$z$ sample, which is the main source of redshifts 
    for our sample, is about 80\% complete at 
    \logms{}$\geq 11.6$ at $0.3 < z < 0.5$. The GAMA survey is 80\% complete down to 
    $10^{10.8}$\msun{} at $z{\sim} 0.3$, but is only 80\% complete to 
    $10^{12.0}$\msun{} at $z{\sim} 0.5$ according to \citet{Taylor2011} (e.g.\ 
    their Fig.~6).
    
    There are 20453 \texttt{S82-MGC} galaxies that are also in the \texttt{hscPho} 
    sample at $0.3 \leq z_{\mathrm{s82}} \leq 0.5$.  
    Because the \texttt{S82-MGC} uses \cmodel{} magnitudes, in this section we use  
    \mcmodel{} for consistency with the \texttt{S82-MGC} catalog.  
    The \texttt{S82-MGC} also uses \texttt{iSEDFit} with similar assumptions as 
    used in this work and we find excellent agreement between HSC \mcmodel{} and 
    the mass derived by \texttt{S82-MGC} which includes NIR data from UKIDSS.
    
    Figure~\ref{fig:mass_complete} compares the number density distributions of 
    galaxies from \texttt{S82-MGC} with those from our sample\footnote{We do not 
    apply any statistical corrections for completeness and hence we do not use the 
    term ``stellar mass function'' to avoid confusion;
    Errors on the distributions are estimated via bootstrap resampling.}. 
    Based on Fig.~\ref{fig:mass_complete}, we conclude that our sample of massive 
    galaxies is reasonably complete down to \logmcmodel{}${\sim} 11.5$ at 
    $0.3 \leq z \leq 0.5$. 
    Given the average difference between \mtot{} and \mcmodel{}, we will focus on 
    galaxies with \logmtot{}$> 11.6$ where our sample shows good completeness. 
    In our discussion section, we will also show results for massive galaxies 
    with $11.4 \le$\logmtot{}$<11.6$ but we caution that our sample is incomplete 
    in this lower mass bin mainly due to the intrinsic incompleteness of the 
    SDSS/BOSS spec-$z$ (see \citealt{Leauthaud2016}).

    

\section{The Final Sample}
    \label{sec:final}
    
	With \mstar{} estimates in hand, we now use the \redm{} galaxy cluster catalog to  
	create a sample of massive central galaxies. 
	The ``central'' galaxy is defined as the galaxy that lives in the 
	center of its own dark matter halo.  
	In contrast, a galaxy in a sub-halo that is orbiting within the virial radius of a 
	more massive halo is referred to as a ``satellite''. 
	To better understand the connection between the structures of massive 
	galaxies and the assembly history of both their stars and dark matter halos, we 
	wish to focus on central galaxies in this work. 
    
\subsection{Candidate massive central galaxies}
    \label{ssec:redmapper}
    
    Although the central galaxies of dark matter halos have unique importance is 
    studying galaxy-halo connection, robust identification of central galaxy is 
    not easy(e.g.\citealt{Yang2007}) in observation.
    First, based on previous results (\citealt{Reid2014, Hoshino2015, Saito2016}),
    the satellite fraction at \logmtot{}$> 11.6$ should become quite low 
    ($\sim 10$\% level). 
    And, we can further use the \redm{} catalog to identify and exclude massive 
    satellites in cluster-level dark matter halos--to further reduce contamination
    from satellite galaxies.
    
    We use \texttt{v5.10} of the \redm{} cluster  catalog
    (e.g.\ \citealt{Rykoff2014}; \citealt{Rozo2015b}). 
    These authors have developed a well-tested red-sequence cluster finder that has 
    been run on SDSS DR8 (\citealt{SDSSDR8}) images. 
    For each cluster, the catalog provides a photometric redshift ($z_{\lambda}$), a 
    cluster richness ($\lambda$), and identifies the most likely central galaxy (the 
    galaxy with the highest $P_{\mathrm{Cen}}$ value). 
    The \redm{}{} catalog also provides a list of member galaxies for each cluster and 
    their associated membership probabilities. 
    Details about the performance of the \redm{}~cluster catalog can be found in 
    \citet{Rozo2014}, \citet{Rozo2015a}, and \citet{Rozo2015b}. 
    
    Several studies have published calibrations between the \redm{}~richness estimate, 
    $\lambda$, and halo mass (e.g.\ \citealt{Saro2015, Farahi2016, Simet2016, 
    Melchior2016}). 
    All these studies consistently find that \redm{} clusters with $\lambda > 20$ 
    generally have $\log (M_{\mathrm{halo}}/M_{\odot}) \geq 14.0$, although the scatter 
    of \mhalo{} at fixed $\lambda$ cannot be ignored.
    Therefore the central galaxies of these \redm{} clusters\footnote{We only use 
    central galaxies with $P_{\mathrm{CEN}} \geq 0.7$} form a sample of massive central 
    galaxies that live in massive halos as described above.
    Such information is also very useful in studying the relationship between galaxy
    structure and environment (Huang\etal in prep.).

    After matching the \texttt{hscZ} sample with the central galaxies of \redm{} 
    clusters with $\lambda \geq 20$ and $P_{\mathrm{Cen}} \geq 0.7$, we find 164 
    matched galaxies at $0.3 \leq z \leq 0.5$.
    This sample of \textbf{central galaxies in more massive halos} will be referred to 
    as the \rbcg{} sample. 
    It is also worth pointing out that, due to the depth and resolution of SDSS 
    images, this \redm{} catalog is not complete down to $\lambda=20$ at 
    $0.3 < z < 0.5$. 
    At $z\geq0.33$, it starts to miss a small fraction of clusters with $\lambda < 30$, 
    but the main results of this work will not be affected by this.
    
    As a next step, we want to construct a sample of central galaxies living in halos 
    with \logmh{}$<14.0$. 
    To achieve this, we identify and remove all galaxies within a cylindrical region 
    around all \redm{} clusters. 
    We use a radius equal to $R_{\mathrm{200b}}$ and the length of the cylinder is 
    set to twice the value of the photometric redshift uncertainty of each cluster.

    We convert $\lambda$ of each cluster to $M_{\mathrm{200b}}$ using the calibration 
    of \citet{Simet2016} and we use the mass-concentration relation from 
    \citet{Diemer2015} to compute $R_{\mathrm{200b}}$. 
    At $0.3 < z < 0.5$, the uncertainty of photo-$z$ is between 0.015 to 0.025, and 
    is enough to exclude cluster members.
    
    After removing galaxies associated with \redm{} clusters, the remaining galaxies 
    in our sample will be dominated by central galaxies living in halos with 
    \logmh{}$< 14.0$. 
    We will refer to this sample of \textbf{galaxies in less massive halos} as the 
    \nbcg{} sample. 
    Because we have rejected satellites in \redm{} clusters, and because above our 
    galaxy mass cut most satellites reside in halos with 
    \logmh$>14.0$ (e.g. \citealt{Reid2014, Hoshino2015, Saito2016, vanUitert2016}), 
    galaxies in the \nbcg{} sample should  be dominated by central galaxies.
    Using the model presented in \citet{Saito2016}, we estimate that in dark matter 
    halos with \logmh$<11.4$, $\sim 7$\% of galaxies with \logmcmodel{}$>11.5$ are 
    satellites.
    This confirms that our sample should have low enough satellite contamination 
    so that it can be taken as representative massive central galaxies.     

\subsection{Summary of Sample Construction}
    \label{ssec:sample}

    Using ${\sim} 100$ deg$^2$ of HSC data, we select a large sample of massive central 
    galaxies with reliable redshift information, and broadly separate them into two 
    categories based on $M_{\mathrm{halo}}$.
    
    The following is a summary of our sample construction. 
        
    \begin{itemize}
        \item \texttt{hscPho} sample: this parent sample consists of bright galaxies 
            with $i_{\mathrm{cModel}} \leq 21.0$, good quality imaging, and reliable 
            \texttt{cModel} photometry in all five HSC bands in the \texttt{S15B} 
            data release. 
            This sample is described in \S \ref{ssec:initial}, and it contains 
            1760845 galaxies. 
        \item \texttt{hscZ} sample: we limit the \texttt{hscPho} sample to galaxies 
            with reliable redshift information. 
            More details of this sample can also be found in \S \ref{ssec:initial}. 
            It provides us 25286 useful galaxies at $0.3<z<0.5$.
        \item With the help of the \redm{} cluster catalog, we further select 
            candidates of massive central galaxies.
            We broadly divide the sample into central galaxies living in halos
            with \logmh{}$\geq 14.0$ (\rbcg{}) and \logmh{}$<14.0$ (\nbcg{}).
            To ensure the sample is \mstar{}-complete and has minimal satellite 
            contamination, we will further focus on the 950 massive galaxies with 
            \logmtot{}$>11.6$ in this work\footnote{As reference, there are 
            2613 massive galaxies with \logmtot{}$>11.5$ in our sample.}. 
           
    \end{itemize}
    
    The division of our sample into two halo mass bins is mainly relevant for Paper II
    (Huang \etal in prep.).  
    For the present paper, we only consider the halo mass dependence on our sample when 
    we evaluate impact of mass estimates on the SMF in \S \ref{ssec:smf}. 
    We show the distributions of redshift, \mtot{} and \minn{} of the massive galaxy 
    sample in Appendix \ref{app:basic}, along with its \mtot{}-$(g-r)$ rest-frame 
    color relation.


  \begin{figure*}
      \centering 
      \includegraphics[width=\textwidth]{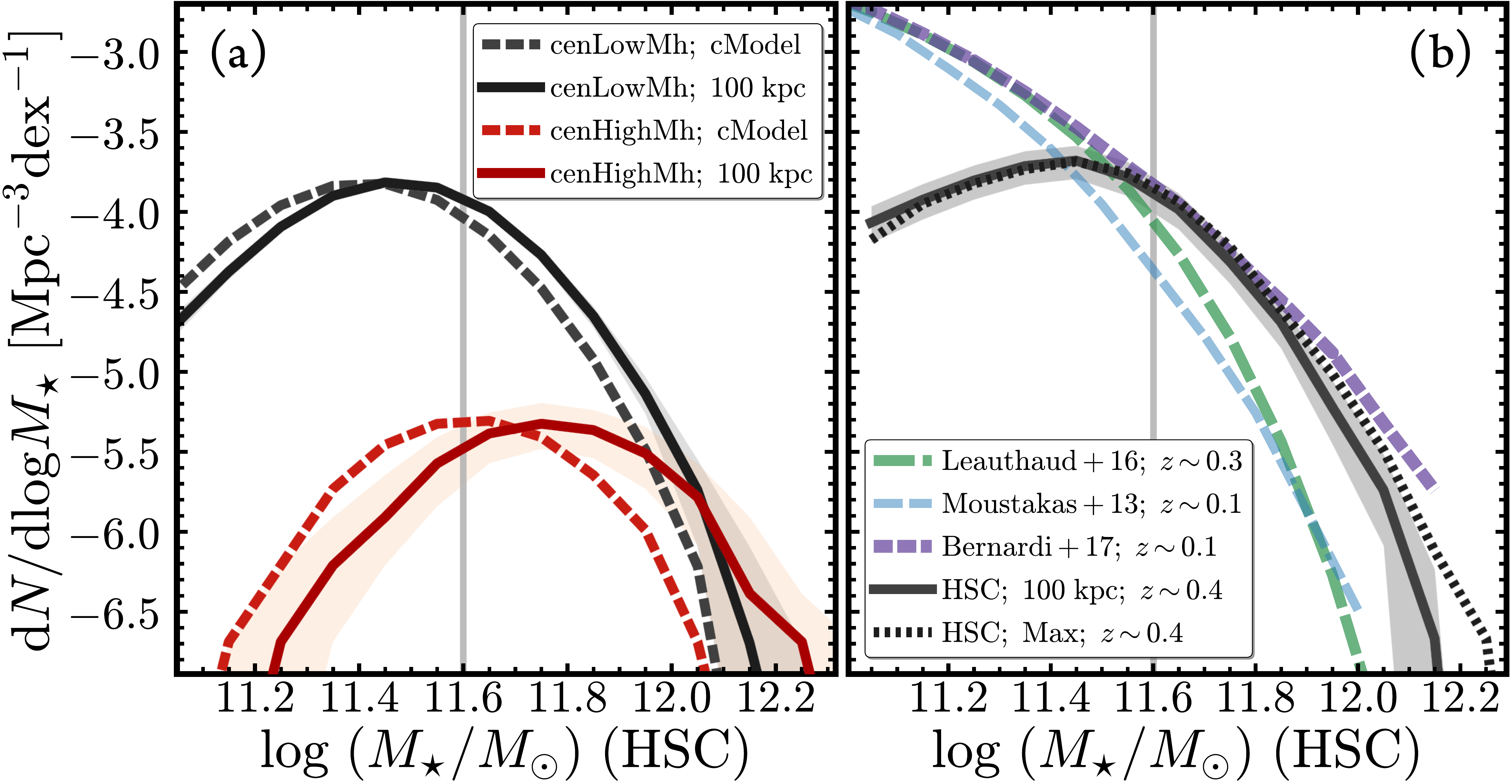}
      \caption{
          \textbf{Left:} Impact of using \mtot{} on the galaxy stellar mass function. 
          Dashed lines correspond to the observed volume density distribution 
          computed using \mcmodel{} whereas solid lines correspond to the distribution
          computed using \mtot{}. 
          We do not apply any completeness correction to the distributions here.
          Here we separate our HSC sample into centrals in halos more massive than
          \logmh{}$\sim14.2$ (red lines) and centrals in halos with \logmh{}$<14.0$ 
          (black lines).
          The impact on the SMF can exceed 0.2 dex for massive central galaxies in 
          very massive halos.
          \textbf{Right}: The \mstar{} volume density distributions of massive 
          HSC galaxies, using both \mtot{} (black solid line) and \mmax{} (black dotted 
          line).  
          Vertical lines on both plots highlight the \logmtot{}$=11.6$ mass limit. 
          The grey shaded region shows the resampling error on the HSC SMF plus an
          additional 20\% uncertainty to account for the fact that we do not include 
          satellite galaxies and that we fail to extract a 1-D profile for $\sim 10$\% 
          of our galaxies. 
          These issues will be addressed in forthcoming work. 
          We compare our results with previous studies: 
          (a): SDSS galaxies at $z{\sim} 0.1$ from \citet{Bernardi2017} with \mstar{} 
          values based on photometry from 2-D \ser{}$+$Exponential model fitting 
          (purple); 
          (b): SDSS galaxies at $z{\sim} 0.1$ from \citet{Moustakas13} based on 
          improved SDSS \texttt{cModel} photometry (blue); 
          (c): \texttt{S82-MGC} galaxies at $0.15 < z< 0.43$ from 
          \citet{Leauthaud2016} based on PSF-matched SDSS-UKIDSS photometry (green).
          }
      \label{fig:smf}
  \end{figure*}

\section{Results}
    \label{sec:result}

\subsection{Impact of Missing Light on the Galaxy Stellar Mass Function}
    \label{ssec:smf}
    
    The Stellar Mass Function (SMF) and its evolution are critical to our understanding 
    of galaxy evolution. 
    On average, the integration of our carefully derived non-parametric \mden{} profiles 
    to 100 kpc (\mtot{}) recovers more light (hence mass) than the \texttt{cModel} method. 
    At \logmtot{} $>11.6$, the difference can sometimes be larger than 0.2 dex. 
    More importantly, the average difference steadily increases with \mtot{}. 
    These differences relate to the intrinsic limitation of the \texttt{cModel} 
    method and also reflect the fact that more massive ETGs tend to have more 
    extended stellar mass distributions (e.g. higher-\ser{} index, 
    \citealt{Graham2003}). Hence, the determination of the ``total luminosities'' of 
    massive galaxies can result in a significant impact on the high mass end of the 
    galaxy stellar mass function (e.g. \citealt{Bernardi2013, DSouza2014, DSouza2015,
    Bernardi2017}).
    
    
    In this paper, we have only measured light profiles for central galaxies. 
    Also, we cannot extract 1-D profiles for $\sim 11$\% of galaxies (because 
    contamination from nearby bright object is too severe; or the its in a late-stage
    on-going merger system). 
    These two effects will lead to a ${\sim}20$\% uncertainty in the amplitude of 
    our volume number density distributions. 
    Both of these effects show dependence on \mtot{} whereas more massive galaxies 
    suffer slightly less from them.
    We are currently working to address these limitations in a forthcoming paper. 
    Our goal in this paper is therefore not to attempt a detailed comparison between 
    different SMFs \citep[e.g.,][]{Bernardi2013, Bernardi2017}. 
    Instead, our goal in this section is simply to characterize the impact of various 
    \mstar{} measurements on the high-mass end of the SMF.   
    Our volume density distributions (will be referred to as SMF for simplicity) 
    turn over at the low-mass end where the sample becomes incomplete, and we do not 
    attempt to apply any completeness correction here.  
    
     Figure \ref{fig:smf} displays the impact of missing light and of different 
     definitions of \mstar{} on the SMF. 
     The left panel of Fig.~\ref{fig:smf} shows the SMF computed using \mcmodel{} 
     and \mtot{} for galaxies that live in low and high mass halos.  
     Figure \ref{fig:smf} shows that the impact of missing flux is more severe for  
     galaxies in high mass halos. 
     In Paper II of this series (Huang\etal in prep.), it will be demonstrated that 
     this occurs because galaxies in more massive halos have more extended stellar 
     envelopes than those in lower mass halos at fixed \mtot{}. 
     Figure \ref{fig:smf} also shows that the use of the HSC \texttt{cModel} 
     magnitude leads to a severe underestimation of \mstar{} at the high mass end 
     of SMF. 
     From the current \texttt{hscPipe}, this problem could relate to the intrinsic 
     limitation of simply model like \cmodel{} and the issue of over-subtracted 
     background. 
     Based on the 1-D surface brightness profiles in $i$-band, background subtraction
     does not seem to be the most serious problem for these $z>0.3$ galaxies 
     (at least not within 100 kpc). 
     And, in the next section, we will show that a significant fraction of these 
     1-D profiles are more extended in the outskirt than the de~Vaucouleurs profile, 
     therefore it is not surprising that \cmodel{} will systematically underestimate 
     their total flux.
     
    
    As discussed in \S~\ref{ssec:mtotal}, even with deep HSC images,
    it is not trivial to clearly define and measure the ``total'' \mstar{} for 
    these massive systems, and the \mden{} profiles of these galaxies often extend 
    well beyond 100 kpc. 
    Is our fiducial 100 kpc radius sufficient to capture most of the flux associated 
    with these galaxies, or does one need to integrate out to even larger radii? 
    To answer this question, we attempt to capture more flux by integrating our 
    \mden{} profiles out to the edge of the image and use the highest \mstar{} 
    achieved during this process as \mmax{}.
    Although the \mden{} profiles become more uncertain at $>100$ kpc, \mmax{} helps 
    us quantify how much extra \mstar{} may be contained at radii greater than 100 kpc.  
    The right hand side of Figure \ref{fig:smf} shows the impact of using \mmax{} 
    instead of  \mtot{} on the SMF.  
    We find that a 100 kpc radius captures a majority of the galaxy mass and that the 
    impact of going from \mtot{} to \mmax{} is relatively small.
    
    In Figure \ref{fig:smf} we also compare our results with the following previous 
    studies:

    \begin{itemize}

        \item The SMF for $0.15 < z < 0.30$ galaxies from the \texttt{S82-MGC} sample
            \citep{Leauthaud2016} where \mstar{} is derived using 
            \texttt{iSEDfit} on PSF-corrected aperture photometry from \texttt{S82}. 
            They adopt the same IMF, dust model, and star-formation model as we do, 
            but use the \texttt{BC03} model instead. 
            Based on the test results in Appendix \ref{app:sed}, we shift this SMF by 
            0.08 dex to add their data to Figure~\ref{fig:smf}.      
       
        \item The SMF derived from a SDSS-\textit{GALEX} sample of $z{\sim} 0.1$ 
            galaxies by \citet{Moustakas13}. 
            Total luminosities in \citet{Moustakas13} are based on SDSS \cmodel{} 
            magnitudes. 
            They use a \m2l{} that is derived via SED fitting using \texttt{iSEDfit} 
            with similar assumptions as the ones adopted here.
            
        \item The observed SMF for $z{\sim} 0.1$ SDSS galaxies from \citet{Bernardi2017}.          
            Total luminosities in \citet{Bernardi2017} are based on 2-D \texttt{SerExp} 
            models (\ser{}$+$ Exponential disk model; integrated to infinity) that 
            recover more flux compared to SDSS \cmodel{} magnitudes 
            (\citealt{Bernardi2013, Meert2015}). 
            They use a \m2l{} that is adopted from the SED fitting results in
            \citet{Mendel2014}. 
            This model also uses the \texttt{FSPS} stellar population model 
            \citep{FSPS}, assumes a \citet{Chabrier2003} IMF, and considers dust 
            extinction using the \citet{Calzetti2000} extinction law (the 
            ``dusty'' model).
            
    \end{itemize}
    
    As mentioned above, our current SMF is subject to an additional 20\% uncertainty 
    that will be addressed in forthcoming work. 
    Hence, we do not attempt a detailed comparison with respect to previous works on 
    the SMF, reserving this for a future study. 
    Here, we simply note that the HSC SMF derived using \mtot{} is close to the one 
    derived by \citet{Bernardi2017} using SDSS data at $z{\sim}0.1$ and assuming 
    the \texttt{SerExp} models.  
    Our sample and the \citet{Bernardi2017} sample have different redshifts. 
    Without a more in-depth study, we cannot ascertain whether or not the apparent 
    agreement between the HSC SMF and the \citet{Bernardi2017} SMF indicates a global 
    lack of evolution in the galaxy SMF, or that other effects are at play 
    (for example, the difference between \mtot{} masses and \texttt{SerExp} masses) 
    that have cancelled redshift evolution.  
    In future work, it will be interesting to perform a more consistent analysis to 
    search for redshift evolution in the galaxy SMF and to study whether or not 
    massive galaxies in HSC are well described by \texttt{SerExp} models.
   
            
   

    

\subsection{Surface Mass Density Profiles}
    \label{ssec:sbp_compare}

  \begin{figure*}
      \centering 
      \includegraphics[width=\textwidth]{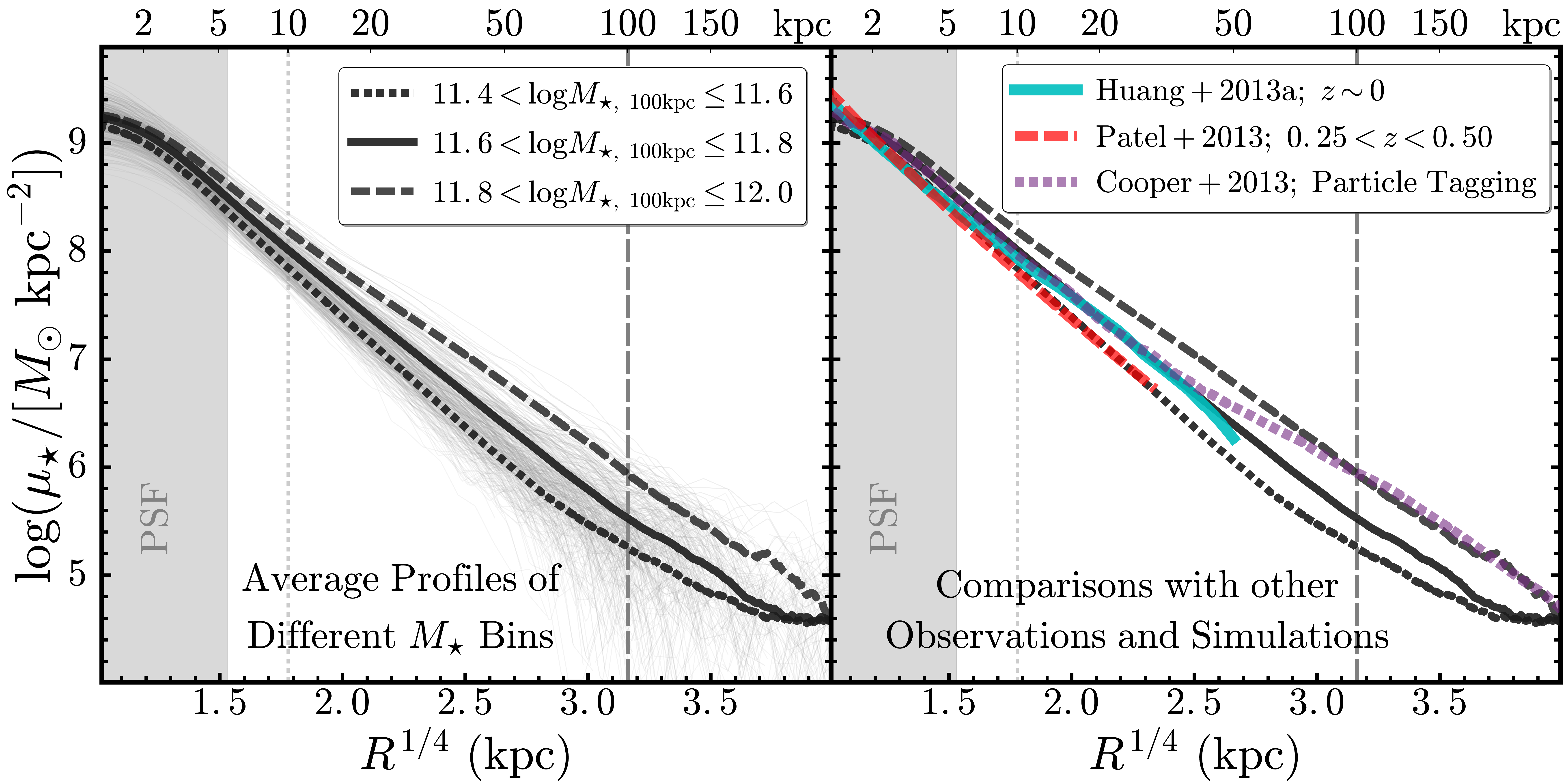}
      \caption{
          \textbf{Left}: Median \mden{} profiles in three total stellar mass bins. 
              Thin grey lines show a random subset of individual profiles.  
              The scatter between the thin grey lines reflects the true scatter in 
              the profiles of massive galaxies (not measurement error). 
              The shaded region highlights the region that is most strongly affected 
              by the seeing. 
              Two vertical lines indicate 10 kpc (thin, dotted line) and
              100 kpc (thick, dashed line). ~~~ 
          \textbf{Right}: comparison between our \mden{} profiles, previous observations, 
              and simulations. 
              The solid cyan line shows the median profile of massive elliptical 
              galaxies at $z{\sim} 0$ from \citet[][]{Huang2013a}. 
              The red long-dashed line shows the median profile of massive galaxies at 
              $0.25 \leq z < 0.50$ observed by \textit{HST} from \citet[][]{Patel2013}. 
              The purple short-dashed line shows the median radial stellar distributions 
              in massive halos from simulation using the particle tagging method
              (\citealt{Cooper2013}).}
      \label{fig:avg_prof}
  \end{figure*}

\subsubsection{General Trends and Comparison with Previous Work}
    \label{sssec:sbp_inter}
          
    Previous work on the structural evolution of massive galaxies has often focused on 
    scaling relations such as the ``\mstar{}-size'' relation.  
    We argue that  comparing \mden{} profiles directly captures more information than
    the \mstar{}-size relation and has the advantage that it bypasses difficult 
    questions about how to accurately define and measure galaxy ``sizes'' and 
    ``masses''.
    
    Fig.~\ref{fig:avg_prof} shows the median \mden{} profiles of massive central 
    galaxies at $0.3 < z < 0.5$ in three \mtot{} bins.
    These median profiles along with their uncertainties are derived using bootstrap
    resampling method. 
    Note that our sample is not complete in the lowest \mtot{} bin, althaough the 
    median \mden{} profile may not be significantly affected. 
    As shown in the left panel of Fig~\ref{fig:avg_prof}, we can comfortably trace
    the \mden{} profiles of these massive galaxies out to 100 kpc \textbf{individually}.  
    At large scales, some of our \mden{} profiles show signs of unphysical truncation and
    fluctuation related to inaccurate sky subtraction. 
    In this paper, we do not use profiles beyond 100 kpc, even though the median 
    \mden{} profiles for the two most massive bins behave reasonably out to 
    ${\sim} 200$ kpc. 
       
    From Figure~\ref{fig:avg_prof} we can see the galaxies in our sample have 
    homogeneous profiles on small radial scales. 
    The amplitude of \mden{} increases with galaxy mass on 10 kpc scales but the slope 
    of \mden{} remains similar.  
    From previous work on this topic, we already know that the inner regions of 
    massive elliptical galaxies display relatively uniform structural (e.g.
    \mden{} profile, isophotal shape: e.g. \citealt{Lauer07, Kormendy2009, 
    Schombert2015}; and kinematic: e.g. \citealt{Cappellari13b}) properties.  
    However, Figure~\ref{fig:avg_prof} reveals a significant diversity in the outer 
    envelopes of massive galaxies.  
    Given the \s2n{} of HSC images at these surface brightness levels, the scatter 
    shown in Figure~\ref{fig:avg_prof} corresponds to \textbf{intrinsic scatter in 
    the stellar envelopes of massive galaxies}. 
    Importantly, Figure~\ref{fig:avg_prof} shows that the global \mden{} profiles 
    of galaxies at these masses are \textbf{clearly not self-similar} out to 100 
    kpc and have outskirts with larger scatter. 
     
    In the right-hand side of Fig.~\ref{fig:avg_prof}, we compare our \mden{} 
    profiles with results from previous work. 
    Deep \mden{} profiles of massive galaxies are rarely available. 
    Even in the nearby universe, it is not trivial to map the low surface
    brightness outskirts of massive galaxies (e.g., 
    \citealt{Capaccioli2015, Iodice2016, Iodice2017, Spavone2017, Mihos2017}). 
    The number of very massive galaxies is also very limited in the local universe.
    For example, according to the \texttt{MASSIVE} survey (\citealt{Ma2014}), there
    are only $\sim 60$-70 massive galaxies with \logms$> 11.6$ (based on K-band 
    luminosity) within 108 Mpc.
       
    Most previous studies have focused on surface brightness profiles instead 
    of mass density profiles.  
    Results can also depend on the stacking technique or the model used to extract 
    the profile (e.g., \citealt{Tal2011, DSouza2015}). 
    \citet{Huang2013a} derived \mden{} profiles for a small sample of very nearby 
    ellipticals (within 100 Mpc; median \logms{} ${\sim} 11.3$) based on relatively 
    shallow images from the Carnegie-Irvine Galaxy Survey 
    (CGS, \citealt{CGS1})\footnote{https://cgs.obs.carnegiescience.edu/CGS/Home.html}.  
    This sample is at very low redshift ($z<0.02$), and so the \mden{} profiles from 
    \citet{Huang2013a} galaxies are accurate to smaller scales (down to $r=1$ kpc) 
    than our HSC ones.  
    Our \mden{} profiles show good agreement with the \citet{Huang2013a} sample in 
    the radial range of overlap (out to 50 kpc). 
    CGS images are deeper than SDSS images in the $r$-band, but the median 
    profiles from \citet{Huang2013a} still only reach to ${\sim} 50$ kpc for 
    $z<0.02$ massive galaxies.
    Meanwhile, our deep HSC images can reliably deliver individual \mden{} profiles 
    for $z{\sim} 0.4$ galaxies out to at least 100 kpc.  
    
    \citet{Patel2013} extracted a median \mden{} profile for massive ETGs at 
    $0.25 < z < 0.50$ using stacked \textit{HST}/ACS images. 
    These galaxies are selected at a constant cumulative number density and are 
    thought to be the progenitors of $z=0$ massive ETGs (e.g. \citealt{Leja2013}).  
    The median \mstar{} of the \citet{Patel2013} sample is 
    ${\sim} 10^{11.2} M_{\odot}$ which is lower than our lowest mass bin. 
    However, \citet{Patel2013} uses the BC03 stellar population model which leads to 
    \mstar{} that are roughly 0.1 dex lower than our FSPS estimates 
    (see Appendix~\ref{app:sed}). 
    Furthermore, the \citet{Patel2013} images are shallower than ours which means 
    that their \mstar{} could still be underestimated due to missing light in the 
    outskirts. 
    Given these two considerations, it is reasonable to roughly compare the 
    \citet{Patel2013} profile with the one in our lowest \mtot{} bin. 
    The superb resolution of the \textit{HST}/ACS images allows \citet{Patel2013} to 
    accurately measure \mden{} profile down to 1 kpc without worrying the smearing 
    effect of seeing. 
    The good agreement between our profiles and the ones derived from HST imaging 
    demonstrates that our profiles are robust at $r\geq 3$ kpc so that we can 
    accurately measure \minn{}.
    
    Finally, we also compare with the predicted median \mden{} profile of central 
    galaxies in massive halos ($13.5 < \log M_{200,c} < 14.0$) from a cosmological 
    simulation where the \mden{} profiles of galaxies are calculated using the 
    particle-tagging technique (e.g., \citealt{Cooper2010}). 
    The simulated \mden{} profile is affected by the resolution limit of the simulation 
    in the inner region, but is in good agreement with our median \mden{} profile for 
    the $11.6 <$ \logmtot{} $< 11.8$ bin within 40 kpc. 
    However, outside 40 kpc, the particle tagging method seems to predict a too 
    prominent stellar halo and has much shallower outer slope compared to our data. 
    In future work, we will compare our HSC profiles with predictions from more 
    advanced hydrodynamic simulations such as \textit{Horizon-AGN} (\citealt{Dubois2014}) 
    and MassiveBlackII (\citealt{Khandai2015}). 
    These data will help fine-tune simulations and to understand the physical mechanisms 
    that drive the assembly of massive galaxies and the build up of stellar halos.

    Table~1 provides tabulated values for the median profiles that are 
    displayed in Fig.~\ref{fig:avg_prof}. 
    These profiles are also available  
    \href{http://www.ucolick.org/\~kbundy/massivegalaxies}{here}. (The files will be made 
    available after the paper is accepted)
    

\subsection{Ellipticity and Color Profiles}
    \label{ssec:ell_color}
    
  \begin{figure*}
      \centering 
      \includegraphics[width=\textwidth]{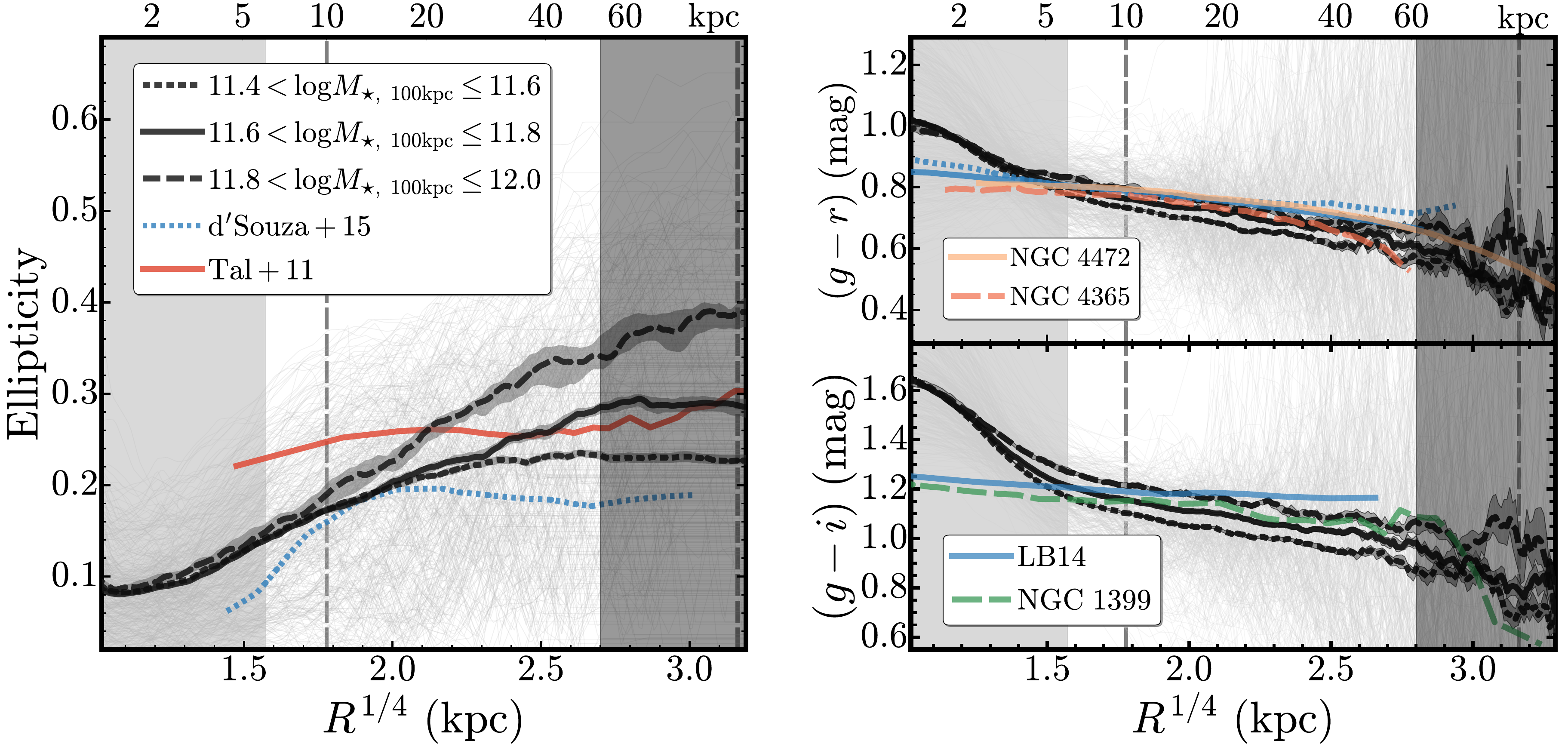}
      \caption{
          Radial profiles of the ellipticty and $k$-corrected rest-frame optical colors 
          of massive galaxies in our sample. 
          The general format of this figure is similar to Fig~\ref{fig:avg_prof}.  
          The \textbf{left panel} displays ellipticity profiles, 
          the \textbf{Upper-right panel} shows $g-r$ color profiles, and  the
          \textbf{lower-right panel} is for $g-i$ color profiles. 
          We compare our results with those from (1) \citet{Tal2011} based on stacking 
          large samples of luminous red galaxies in SDSS at $z{\sim} 0.4$ 
          (solid red line on the left panel), 
          (2) the results from a stacking analysis of nearby massive 
          galaxies with high concentration index ($C>2.6$) in 
          \citet[][blue dash lines on the left and upper-right panels]{DSouza2014}, 
          (3) and the average $g-r$ and $g-i$ color profiles 
          from a large sample of nearby elliptical galaxies in \citet[][blue, solid 
          lines on both right panels]{LaBarbera2010}.
          For color profiles, we also compare with deep observations of a few nearby 
          massive ETGs: the $g-r$ color profiles of NGC 4472 (\citealt{Mihos2013}) 
          and NGC 4365 (\citealt{Mihos2017}), and the $g-i$ profile of NGC 1399 
          (\citealt{Iodice2016}).
          }
      \label{fig:ell_color}
  \end{figure*}
    
    So far, we have focused on 1-D \mden{} profiles. 
    We now consider ellipticity and $k$-corrected optical color profiles.
	We extract ellipticity profiles using \texttt{Ellipse} by leaving the shape of 
	each isophote as a free parameter. 
	We also apply the isophotal information derived in the $i$-band directly to other 
	filters and extract 1-D $g-r$ and $g-i$ color profiles.
	We apply the Galactic extinction correction and the \texttt{iSEDFit} $k$-correction 
	to the color profiles. 
	Smearing effect of seeing will make the central isophotal shape rounder than the 
	real value, while seeing differences between filters will bias the central 
	color\footnote{Since the HSC $i$-band always has better seeing, the central color 
	will become redder is seeing effects are not accounted for}. 
	On large scales, it is more difficult to extract reliable ellipticity and 
	color profiles out to 100 kpc: at low surface brightness levels, the isophotal 
	shape becomes unstable and is easily affected by contamination. 
	Color profiles are also more difficult to extract out to large radial scales
	because  getting the color accurately depends on the background subtraction and 
	the noise levels in both bands. 
	In this paper, we will focus on the median ellipticity and color profiles between 
	8--60 kpc where we can safely ignore the issues described above. 
	
	The ellipticity of the isophotes contains information about the 3-D geometry 
	(e.g. \citealt{Tremblay1995, Tremblay1996, Chang2013, RodriguezPadilla2013, 
	Mitsuda2017}) and kinematics (e.g. \citealt{Cappellari2012, Weijmans2014}) of 
	stars in massive galaxies.  
	The left panel of Fig \ref{fig:ell_color} shows the ellipticity profiles of 
	massive galaxies and highlights the median profiles for the same three \mtot{} 
	bins as used in Fig \ref{fig:avg_prof}.  
	Our results are also compared with previous work based on image stacking techniques 
	(the PSF-removed $i$-band results from \citealt{Tal2011}, and
    concentrated galaxies with $11.0<$\logms{}$<11.4$ from \citealt{DSouza2015}).
	As expected, ellipticity profiles from image stacking methods yield results that 
	are more shallow than when ellipticity can be measured on a galaxy-by-galaxy basis.     
	Uncertainties in how to align galaxies, and the intrinsic isophotal twist can 
	lead to this effect. 

	In general, we find that the ellipticities of massive galaxies slowly increase 
	with radius. 
	This trend can even be seen directly "by eye" in HSC images (e.g. 
	Fig \ref{fig:sdss_compare}). 
	More interestingly, the ellipticity profiles vary with \mtot{}: 
	at ${\sim} 10$ kpc, the median ellipticity ($< 0.2$) is similar for all three 
	redshift bins, but the ellipticity of the outer stellar halo increases with 
	\mtot{}. 
	Galaxies with \logmtot{}$>11.8$ have median ellipticity profiles that become 
	steeper at $>10$ kpc. 
	The ellipticity of the outer profile steadily increases from 
	$e\le 0.2$ to $e{\sim} 0.4$ at 50-60 kpc.
    
	This is consistent with studies of nearby massive galaxies, both from 1-D 
	ellipticity profiles and 2-D modeling results \citep{Porter1991, Gonzalez2005, 
	Zibetti2005, Spavone2017, Huang2013a, Oh2017}. 
	However, to the best of our knowledge, our HSC results are the first to show 
	clear evidence that:  
	a) the ellipticity of stellar halo in massive ETGs depends strongly on 
	\mtot{}, and 
	b) the ellipticity of stellar halo also relates to the slope of the \mden{}
	profile (see Fig~\ref{fig:avg_prof} and Fig~\ref{fig:ell_color}; 
	we will discuss this more in Paper II). 
    
	If accreted stars dominate the stellar halos of massive galaxies, this 
	mass-dependent ellipticity profile may contain clues about the assembly history 
	of massive galaxies (e.g. average time since last merger, average merger 
	mass ratio).
	Previous simulations often focused on reproducing the rounder average shape 
	and slow-rotating nature of massive ETGs (e.g. \citealt{Wu2014}), and the 
	projected ellipticity profiles and their correlations with the kinematic of 
	stars and merging history have not been carefully explored. 
   
    Regarding the color profiles, it is well known that massive elliptical galaxies 
    have shallow and negative color gradients that reflect radial variations in their  
    stellar populations (e.g.\ \citealt{Carollo1993, LaBarbera2012}) and hence contain
    information about the assembly history of their stellar halos (e.g. 
    \citealt{Hirschmann2015}). 
    The right panels of Fig \ref{fig:ell_color} show the $k$-corrected $(g-r)$ and 
    $(g-i)$ color profiles for all galaxies in our sample, together with the median 
    profiles in three \mtot{} bins. 
    We find the the median rest-frame $(g-r)$ and $(g-i)$ color decreases at 
    larger radii, but there does not appear to be a significant \mstar{} dependence 
    in the gradient of rest-frame optical colors. 
    We also compare our results with the stacked color profiles from  
    \citealt{LaBarbera2010}\footnote{We use the median color profiles of high-mass 
    ETGs; the original profile is in units of $R_{\mathrm{e}}$, we use a typical 
    $R_{\mathrm{e}}=8.0$ kpc to convert it into physical kpc.} and 
    \citealt{DSouza2014}.
    The median color profiles from HSC images are systematically steeper than the 
    stacked SDSS ones. 
    Considering differences in the response curves between HSC and SDSS filters, 
    together with the uncertainties of color measurements, the HSC $(g-r)$ color 
    profiles are in fairly good agreement with those from SDSS. 
    However, the $(g-i)$ profiles of HSC galaxies are steeper compared to SDSS. 
    The SDSS $i$-band suffers from the so-called ``red-halo'' effect
    (e.g. \citealt{Wu2005}, \citealt{Tal2011}).  
    This is due to the fact that the SDSS $i$-band PSF has more prominent wing 
    than other bands. 
    Because the PSF model does not capture these wings, this artificially distributes 
    more flux to the outskirts, and leads to apparently redder colors in the low 
    surface brightness outskirts of galaxies. 
    Because HSC uses thick CCDs, HSC $i$-band images do not suffer from this effect 
    and can be used to determine galaxy colors with higher accuracy. 
        
    
    Fairly steep color profiles have been observed in several very nearby massive 
    ETGs.  
    Fig \ref{fig:ell_color} shows the $g-r$ color profiles of NGC 4472 
    (\citealt{Mihos2013}) and NGC 4365 (\citealt{Mihos2017})\footnote{
    Both are converted from ($B-V$) colors.}, and the $g-i$ profile of NGC 1399 
    (\citealt{Iodice2016}). 
    These individual profiles display similar color gradients as our HSC sample.   

  \begin{figure*}
      \centering 
      \includegraphics[width=\textwidth]{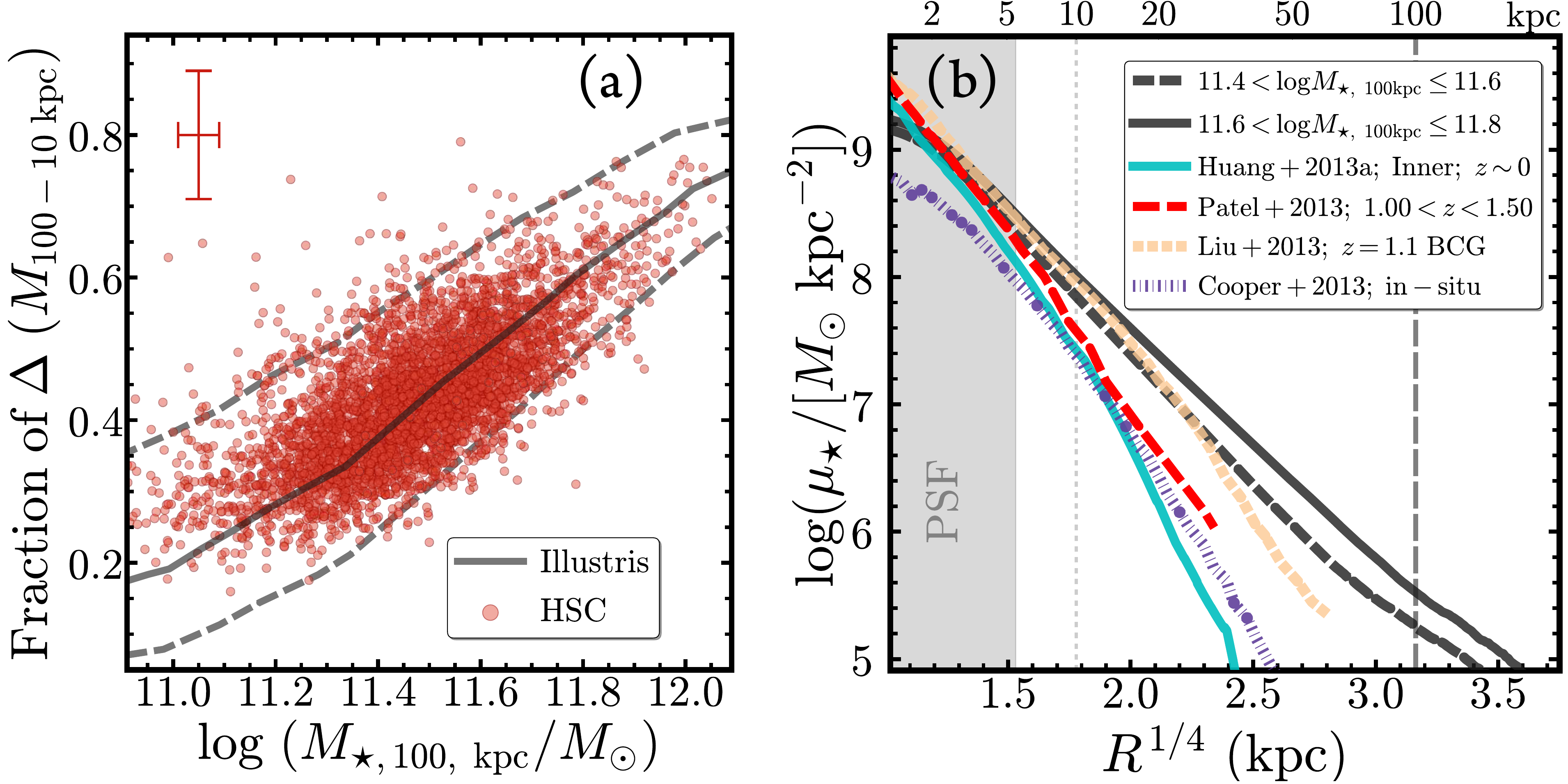}
      \caption{
          \textbf{Left:} Ratio of the fraction of stars between 10 to 100 kpc to the 
          total galaxy mass \mtot{}. 
          We adopt this ratio as a proxy for the fraction of ex-situ stars in our 
          sample. 
          Typical observational uncertainties are shown in the upper left hand corner. 
          The solid grey line shows the predicted relation  derived from the Illustris
          simulation at $z=0$ (Fig 4 in \citealt{RodriguezGomez2016}). 
          Regions between the grey dashed lines correspond to the range between the 
          16 and 84 percentile of the distribution. 
          \textbf{Right:} comparison between our median \mden{} profiles with the 
          inner component of the structural decomposition of massive elliptical 
          galaxies at $z<0.02$ from \citet[][Cyan, solid]{Huang2013a}.
          At higher redshifts, the \mden{} profiles of massive galaxies should be 
          dominated by the in-situ component. 
          We compare our profiles with the median \mden{} profile of massive galaxies 
          at $1.0 < z < 1.5$ from \textit{HST} observations 
          \citet[][Red, dashed]{Patel2013}. 
          Both these comparisons suggest that the \mden{} profile within 10 kpc is 
          dominated by \textit{in situ} stars, but there are already contributions
          from the accreted stars at very high \mstar{} end. 
          We also compare with the \mden{} profile of a very massive cD galaxy at
          $z{\sim} 1.1$ discovered by \citet[][Yellow, dashed]{Liu2013} in the Hubble 
          Ultra-Deep Field. 
          It is likely that this object will grow into one of the very massive central
          galaxy in our sample.
          It is interesting to see that its \mden{} profile is very similar to the 
          HSC one of the most massive \mtot{} bin in the inner $\sim 20$ kpc, so the 
          following growth should mostly happen in the outskirt. 
          }
      \label{fig:discussion_1}
  \end{figure*}

\section{Discussion}
    \label{sec:discussion}
    
    In this paper, we have used data from the HSC survey that is both simultaneously 
    deep and wide to trace the stellar mass distributions of $0.3 < z < 0.5$ massive 
    galaxies out to $>100$ kpc and to reveal the mass-dependent nature of their 
    stellar halos.
    Here we briefly discuss the scientific implications of our results.
    

\subsection{The Formation of Massive Galaxies and the Assembly of Their Outer Halos}
    \label{ssec:twophase}
            
    We find that the outer halos of massive elliptical galaxies grow more prominent 
    and more elliptical with increasing stellar mass.
    According to the two-phase formation scenario, the inner 5--10 kpc of these 
    massive central galaxies are formed at $z>1$ during an intense period of 
    \textit{in situ} star formation. 
    The outskirts of massive galaxies are then built up through a more gradual second 
    phase of evolution (the \textit{ex situ} phase) that is dominated by mass assembly via 
    accretions.
    Non-dissipative mergers, especially minor mergers\footnote{Normally minor merger
    means the one with stellar mass ratio smaller than 1:3 or 1:4}, mostly deposit stars
    in the outskirts of centrals and do not have a large impact on the central \mden{} 
    profile (e.g. \citealt{BoylanKolchin2008, Oogi2013, Bedorf2013}).
    Given the stochastic nature of the merging process, it is easy to understand why 
    the \mden{} profiles of massive galaxies are similar in the inner region but show
    a large scatter in the outer region. 
        
    State of the art hydrodynamic simulations of massive galaxy formation predict 
    that the fraction of accreted stars should strongly increase with stellar 
    mass and that for very massive galaxies, the \textit{ex-situ} may reach up to 
    50-90\% of the total galaxy mass (e.g.
    \citealt{Oser2010, Cooper2013, Dubois2013, LeeYi2013, Hirschmann2015,
    RodriguezGomez2016}). 
    
    This picture is supported by our observations that massive galaxies display more 
    prominent stellar halos as well as by their  negative color gradients. 
    The fact that the outskirts of these galaxies are slightly bluer than the inner
    regions, is consistent with the picture that stellar halos are built up by 
    series of minor mergers (average merger mass-ratio between 1:3-1:10; e.g. 
    \citealt{Huang2016}) as less massive ETGs are typically bluer. 

    Also according to this picture, the shape of the stellar halo should preserve
    information about the merging history and possibly even about the shape of the 
    dark matter halo. 
    Simulations show the shape of the stellar and dark matter halos are closely 
    correlated (e.g. \citealt{Wu2014}) for slowly-rotating massive ETGs having 
    undergone multiple minor-mergers.
    The more elongated outer halo and the trend between the ellipticity profile and 
    stellar mass may reflect the orbital properties of accreted satellite. 
    In simulations, satellite orbits become more radial for more massive halos 
    (e.g. \citealt{Murante2007, Wetzel2011, Jiang2015}. 
    Satellites infalling along radial orbits may help to form elongated stellar 
    halos.
    In principle, these results may explain the trend that we see. 
    However, in contrast, simulation from \citet{Wu2014} predict that more massive 
    galaxies should have \emph{rounder} outskirts which is the opposite trend 
    compared to HSC. 
    This difference warrants further investigation.
    
    At larger scales, the distribution of satellites galaxies in massive halos is 
    found to be aligned with the major axis of the central galaxy 
    (e.g. \citealt{Brainerd2005, Yang2006b, NiedersteOstholt2010, 
    HuangMandelbaum2016}). 
    This alignment signal is found to be stronger for more luminous galaxies 
    living in more massive halos (e.g. \citealt{Hirata2007}).
    The fact that more massive central galaxies have steeper ellipticity profiles 
    and become increasingly more elongated in the outskirt may arise because of 
    such alignment. 
    Moreover, the shape of the dark matter halo can be estimated by analyzing 
    satellite distributions and weak lensing profiles (\citealt{ClampittJain2016}). 
    It is interesting to point out that the most recent measurement by 
    \citet{Shin2017} around SDSS clusters show a halo axis-ratio of ${\sim}0.55$, 
    which is only slightly more elongated than the stellar halo of galaxies in our 
    highest \mtot{} bin.
    
    

\subsection{Aperture Masses as Proxies of the \textit{In situ} and Accreted Stars}
    \label{ssec:insitu}
    
    Recent hydrodynamic simulations of galaxy formation often separate stars in 
    galaxies into \textit{in situ} and \textit{ex situ} components. 
    In simulations, the \textit{in situ} component is often defined as the stars 
    formed in the halo of the ``main progenitor''.  
    The spatial distribution, kinematic, and stellar population properties of both 
    components are important theoretical predictions. 
    Among various predictions, the faction of \textit{ex situ} stars and the scaling 
    of the \textit{ex situ} fraction with stellar mass is perhaps the first aspect to 
    test. 
    However, on the observational side, there is no straightforward way to actually
    disentangle these two components for massive galaxies. 
    Disk galaxies are more easily decomposed into multiple components and recently, 
    deep surveys of nearby disk galaxies have started to provide constraints on 
    their stellar halos (e.g. \citealt{Courteau2011, Merritt2016, 
    Harmsen2017})\footnote{Although it is still not certain that all stellar halos
    around disk galaxies are made out of \textit{ex situ} stars.}.
    
    For more massive ETGs, results so far mainly depend on multi-component model 
    fitting and image stacking analyses (e.g. \citealt{Huang2013a, DSouza2014,
    Spavone2017}).
    In this work, we propose that \mstar{} computed within different fixed physical
    elliptical apertures are worth exploring as proxies of the \textit{ex situ} 
    fraction. 
    We propose to use the mass within 10 kpc (\minn{}) and 100 kpc (\mtot{}) as 
    proxies for the \textit{in situ} component and for the total \mstar{}. 
    
    On the left panel of Fig \ref{fig:discussion_1} we show the relation between 
    \mtot{} and the fraction of mass between 10 and 100 kpc (as a proxy of the mass of 
    the accreted component).
    We compare this with the fraction of \textit{ex-situ} stars predicted by the 
    Illustris simulation (\citealt{RodriguezGomez2016}).  
    We find that our proxy for the accreted mass component correlates strongly with  
    \mtot{} and that this relation is surprisingly consistent with predictions from
    \citet{RodriguezGomez2016}.  
    Given the limitations of the Illustris simulation, and the imperfect nature of 
    our \textit{ex situ} fraction proxy, the almost perfect agreement seen in 
    Figure \ref{fig:discussion_1} may well be a coincidence.  
    However, it is encouraging to see that our simple proxy for the fraction of 
    \textit{ex situ} stars seems to match both the slope and the scatter of the 
    predicted relation.  
    There exists other hydrodynamic simulations that predict significantly different 
    \textit{ex situ} fractions compared to Illustris (e.g. \citealt{Lackner2012, 
    Qu2017}). 
    In future work, we will explore more detailed comparisons between our data and
    predictions from hydrodynamic simulations and we will investigate to what degree 
    simple elliptical aperture masses may be used to trace \textit{ex situ} fractions. 
    
   
    How well justified is our choice of \minn{} as a proxy of the \textit{in-situ} 
    component?  
    A full investigation is beyond the scope of this paper. 
    Here we simply present comparisons between our profiles and  
    a) several estimates of the \textit{in-situ} component and 
    b) observations of high redshift massive galaxies that should be dominated by 
    the \textit{in situ} component.  
    In particular, we compare with:
  
    \begin{enumerate}
        
        \item The median \mden{} profiles of massive ETGs at $1.0 < z < 1.5$ from
            \citealt{Patel2013}.
            These are considered to be the progenitors of 
            ${\sim} 10^{11.5} M_{\odot}$ ETGs at $z=0$ and their inner region 
            should be dominated by \textit{in situ} stars. 
    
        \item The inner component of $z{\sim} 0$ ellipticals from the 2-D 
            decomposition of \citet{Huang2013a}. 
            \citet{Huang2013b} show that this inner component is structurally 
            similar to the compact ``red nuggets'' at high-$z$. 
            
        \item The \textit{in situ} components of simulated central galaxies in 
            massive halos from \citet{Cooper2013} (the inner ${\sim} 5$ kpc is 
            quite uncertain due to the resolution).  
            These \mden{} profiles are generated using the particle tagging 
            method (see \citealt{Cooper2010})
    
    \end{enumerate}

    We also compare with a uniquely massive BCG at high redshift: 
    a \logms{}${\sim} 10^{11.4} M_{\odot}$ BCG with a distinctive ``cD''-like envelope 
    at $z{\sim} 1.1$ (\citealt{Liu2013}).  
    This high redshift galaxy has a \mden{} profile that follows the median \mden{} 
    profile of our $11.6\leq$\logmtot{}$<11.8$ sample nicely at $\mathrm{R} < 20$ 
    kpc, but the profile becomes much steeper in the outskirt. 
    This suggests that the inner ``core'' of massive BCGs are already in pace at 
    $z{\sim} 1$ while the outer halo is still being assembled. 

    These simple comparisons certainly support the idea that \minn{} mainly consists 
    of \textit{in situ} stars whereas mass at $\mathrm{R} > 15$-20 kpc is dominated 
    by \textit{ex situ} component.  
    Meanwhile, this comparison also shows that the \texttt{in situ} component may 
    extend beyond 10 kpc\footnote{We convert these \mden{} profiles to the same 
    \citealt{Chabrier2003} IMF; but there are still differences in median \mstar{} 
    and details in the \m2l{} estimates}. 
    Further comparisons with massive galaxies at high redshift and with 
    hydrodynamical simulations will help understand which radius is optimal for 
    probing the \textit{in situ} mass.
      
    


\section{Summary and Conclusions}
    \label{sec:summary}
    
    In this work, we study how the projected stellar mass density profiles and other 
    structural properties of massive central galaxies depend on their total stellar 
    mass using deep images from the Subaru HSC survey. 
    With the help of this high-quality and wide area data set, we directly map the 
    stellar mass distributions of ${\sim}7000$ massive central galaxies at 
    $0.3 < z < 0.5$ out to $>100$ kpc without resorting to stacking techniques. 
    We group massive central galaxies into two categories based on their host halo 
    mass (\mhalo{}$\simgt 10^{14.0} M_{\odot}$ and \mhalo{}$\simlt 10^{14} M_{\odot}$)
    and three bins of \mtot{}.  
    Our main results are:  
    
    \begin{enumerate}
        
        \item We find that the ``total'' \mstar{} of these massive galaxies can be 
            significantly underestimated with shallow imaging data such as SDSS and/or 
            oversimplified model assumptions (e.g. the \texttt{cModel} or single-\ser). 
            In contrast to previous work, our results do not depend on stacking or any 
            parametric models. 
            Moreover, the degree to which stellar mass is underestimated depends on 
            \mtot{}. 
            Simple model misses more light for massive galaxies because they have 
            more extended envelopes. 
            There is also a \mhalo{}-dependence of this effect and naive luminosity
            estimates will miss more light for BCGs in more massive halos compared 
            to centrals in less massive halos.
            These effects need to be carefully taken into account when discussing 
            topics such as the evolution of the galaxy stellar mass function.     
        
        \item We show that the \mden{} profiles of massive galaxies are relatively
            homogenous within 10-20 kpc. 
            However, there is large scatter in outer profiles of massive galaxies. 
            Galaxies with higher \mtot{} show more prominent stellar halos and have 
            shallower outer \mden{} slopes. 
            Assuming that stellar halos are dominated by accreted stars, this result 
            is consistent with the two-phase formation picture of massive galaxies.
             
        \item We show that, on average, massive galaxies have positive ellipticity 
            gradients out to at least 60 kpc. 
            The average ellipticity profile also depend on \mtot{}: more massive 
            galaxies tend to have steeper ellipticity gradients and become more 
            elongated in stellar halos. 
            On the other hand, the average $(g-r)$ and $(g-i)$ color gradients do 
            not show clear dependence on \mtot{} within 10-60 kpc.      
            
    \end{enumerate}

    These results highlight the advantages of wide area, deep, and high-quality imaging 
    for studying the evolution of massive galaxies. 
    At present, the HSC survey has already doubled its sky coverage to 
    ${\sim} 200$ deg$^2$, and provides a much larger sample of massive central galaxies. 
    In the near future, we will extend this work to lower \mtot{} by using photometric 
    redshifts, and we will also apply 2-D photometric methods (e.g.\ \citealt{Huang2013a}) 
    to take advantage of the multi-wavelength nature of the HSC survey 
    (e.g. \citealt{Huang2016}). 
    Our current work can also be combined with weak lensing measurements of the dark 
    matter halos of massive galaxies and physical insights into the assembly histories 
    of these galaxies can be gained by comparing with cosmological hydro-simulations 
    such as Illustris (\citealt{Vogelsberger2014}, \citealt{Genel2014}), 
    EAGLE (\citealt{Schaye2015}, \citealt{Crain2015}), or \textit{Horizon-AGN} 
    (\citealt{Dubois2014}).

  
\section*{Acknowledgements}

  The authors thank Rachel Mandelbaum and Frank van~den~Bosch for insightful 
  discussions and comments; 
  thank Shun Saito for helping us estimate the fraction of satellite 
  galaxies in our sample;
  thank Feng-Shan Liu for sharing the \mden{} profile of the $z\sim1$ BCG from 
  his work.

  The Hyper Suprime-Cam (HSC) collaboration includes the astronomical communities of 
  Japan and Taiwan, and Princeton University.  The HSC instrumentation and software were
  developed by the National Astronomical Observatory of Japan (NAOJ), the Kavli Institute
  for the Physics and Mathematics of the Universe (Kavli IPMU), the University of Tokyo,
  the High Energy Accelerator Research Organization (KEK), the Academia Sinica Institute
  for Astronomy and Astrophysics in Taiwan (ASIAA), and Princeton University.  
  Funding was contributed by the FIRST program from Japanese Cabinet Office, the Ministry 
  of Education, Culture, Sports, Science and Technology (MEXT), the Japan Society for 
  the Promotion of Science (JSPS), Japan Science and Technology Agency (JST), the
  Toray Science Foundation, NAOJ, Kavli IPMU, KEK, ASIAA, and Princeton University.
   
  Funding for SDSS-III has been provided by the Alfred P. Sloan Foundation, the
  Participating Institutions, the National Science Foundation, and the U.S.  Department of
  Energy. The SDSS-III web site is http://www.sdss3.org.  SDSS-III is managed by the
  Astrophysical Research Consortium for the Participating Institutions of the SDSS-III
  Collaboration including the University of Arizona, the Brazilian Participation Group,
  Brookhaven National Laboratory, University of Cambridge, University of Florida, the
  French Participation Group, the German Participation Group, the Instituto de Astrofisica
  de Canarias, the Michigan State/Notre Dame/JINA Participation Group, Johns Hopkins
  University, Lawrence Berkeley National Laboratory, Max Planck Institute for
  Astrophysics, New Mexico State University, New York University, Ohio State University,
  Pennsylvania State University, University of Portsmouth, Princeton University, the
  Spanish Participation Group, University of Tokyo, University of Utah, Vanderbilt
  University, University of Virginia, University of Washington, and Yale University.
  
  The Pan-STARRS1 Surveys (PS1) have been made possible through contributions of the 
  Institute for Astronomy, the University of Hawaii, the Pan-STARRS Project Office, 
  the Max-Planck Society and its participating institutes, the Max Planck Institute 
  for Astronomy, Heidelberg and the Max Planck Institute for Extraterrestrial Physics, 
  Garching, The Johns Hopkins University, Durham University, the University of Edinburgh, 
  Queen's University Belfast, the Harvard-Smithsonian Center for Astrophysics, the Las 
  Cumbres Observatory Global Telescope Network Incorporated, the National Central 
  University of Taiwan, the Space Telescope Science Institute, the National Aeronautics 
  and Space Administration under Grant No. NNX08AR22G issued through the Planetary 
  Science Division of the NASA Science Mission Directorate, the National Science 
  Foundation under Grant No. AST-1238877, the University of Maryland, and Eotvos 
  Lorand University (ELTE).
  
  This paper makes use of software developed for the Large Synoptic Survey 
  Telescope. We thank the LSST Project for making their code available as free 
  software at http://dm.lsstcorp.org.
  
  This research was supported in part by the National Science Foundation under Grant 
  No. NSF PHY11-25915. 
 
  This research made use of:
  \href{http://www.stsci.edu/institute/software_hardware/pyraf/stsci\_python}{\texttt{STSCI\_PYTHON}},
      a general astronomical data analysis infrastructure in Python. 
      \texttt{STSCI\_PYTHON} is a product of the Space Telescope Science Institute, 
      which is operated by AURA for NASA;
  \href{http://www.scipy.org/}{\texttt{SciPy}},
      an open source scientific tools for Python (\citealt{SciPy});
  \href{http://www.numpy.org/}{\texttt{NumPy}}, 
      a fundamental package for scientific computing with Python (\citealt{NumPy});
  \href{http://matplotlib.org/}{\texttt{Matplotlib}}, 
      a 2-D plotting library for Python (\citealt{Matplotlib});
  \href{http://www.astropy.org/}{\texttt{Astropy}}, a community-developed 
      core Python package for Astronomy (\citealt{AstroPy}); 
  \href{http://scikit-learn.org/stable/index.html}{\texttt{scikit-learn}},
      a machine-learning library in Python (\citealt{scikit-learn}); 
  \href{http://www.astroml.org/}{\texttt{astroML}}, 
      a machine learning library for astrophysics (\citealt{astroML});
  \href{https://ipython.org}{\texttt{IPython}}, 
      an interactive computing system for Python (\citealt{IPython});
  \href{https://github.com/kbarbary/sep}{\texttt{sep}} 
      Source Extraction and Photometry in Python (\citealt{PythonSEP});
  \href{https://jiffyclub.github.io/palettable/}{\texttt{palettable}},
      color palettes for Python;
  \href{http://dan.iel.fm/emcee/current/}{\texttt{emcee}}, 
      Seriously Kick-Ass MCMC in Python;
  \href{http://bdiemer.bitbucket.org/}{\texttt{Colossus}}, 
      COsmology, haLO and large-Scale StrUcture toolS (\citealt{Colossus}).


\bibliographystyle{mnras}
\bibliography{redbcg}


\appendix
 
  \begin{figure}
      \centering 
      \includegraphics[width=\columnwidth]{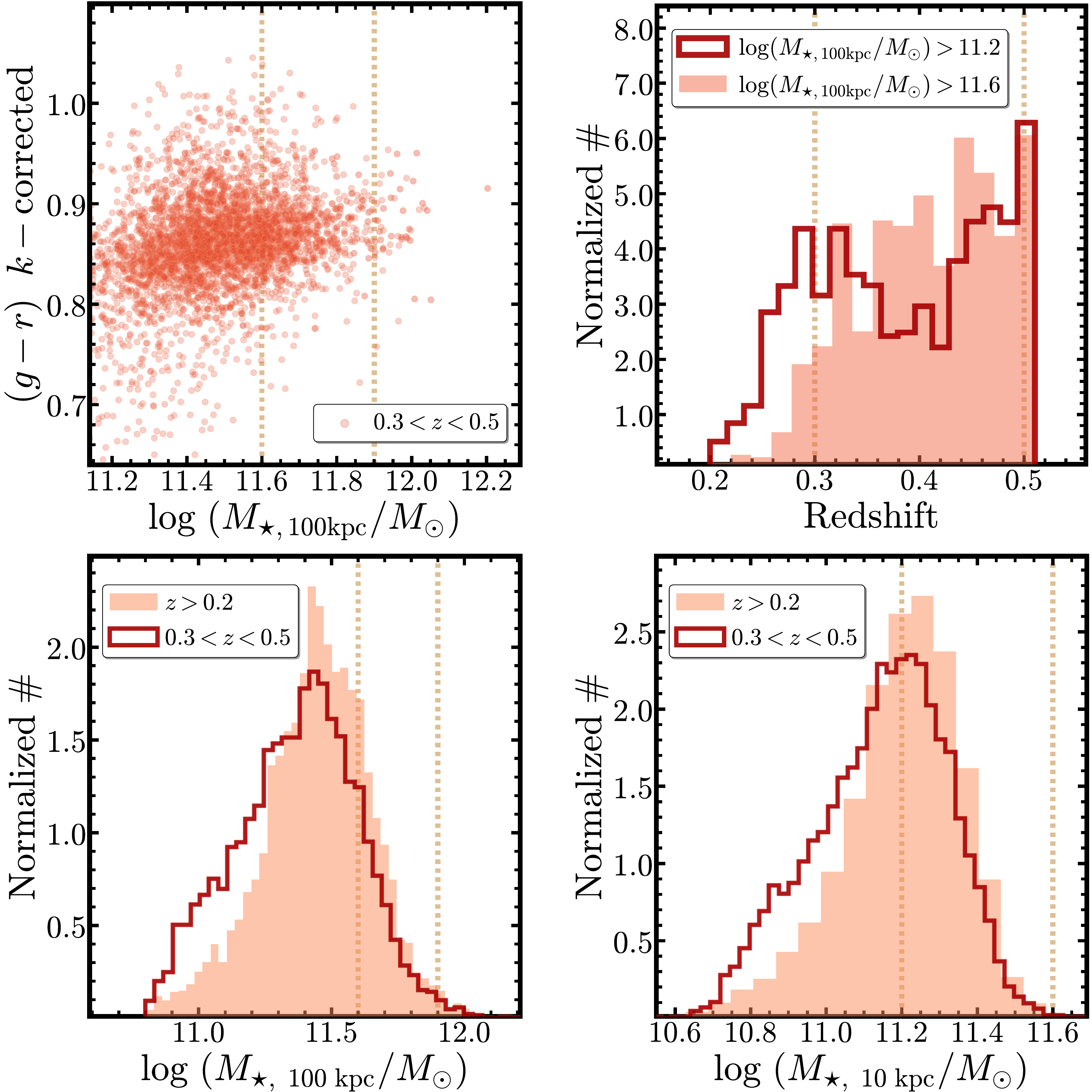}
      \caption{
          \textbf{Top-left}: The \logmtot{}-$(g-r)$ rest-frame color relation for the 
          HSC massive galaxies.
          We $k$-correct the color using the \texttt{iSEDFit} fitting results.
          Massive galaxies form a ``red-sequence'' on this figure, and there is 
          little contamination from blue object at high-mass end.~~          
          \textbf{Top-right}: the redshift distribution of the massive galaxies. 
          The filled and empty histograms are for the \logmtot{}$>11.6$ and 
          \logmtot{}$>11.2$ galaxies.
          The vertical lines highlights the $0.3\leq z \leq 0.5$ redshift range.~~
          \textbf{Bottom-left}: the distributions of \mtot{} of massive galaxies in 
          this sample. 
          Filled histogram shows the distribution for $0.3 < z < 0.5$ galaxies 
          used in this work. 
          And the empty histogram shows the distribution for the whole $z>0.2$ sample
          as comparison.~~
          \textbf{Bottom-right}: the distributions of \minn{} in similar format. 
      }
      \label{fig:sample_stats}
  \end{figure}
   
\section{A. Basic Statistical Properties of the Sample} 
	\label{app:basic} 
    
    Here we show the basic statistics of the massive galaxies used in this work.
    On the top-left panel of Fig~\ref{fig:sample_stats}, we show the \mtot{}-color 
    relation using the $k$-corrected rest-frame $(g-r)$ color. 
    These massive galaxies form a clear ``red-sequence'' with little contamination 
    from the ``blue cloud'' at the very high-mass end.
    
    In the rest of Fig~\ref{fig:sample_stats}, we also show the distributions of 
    redshift, \mtot{}, and \minn{}.
    In this work, we focus on the massive galaxies with \logmtot{}$>11.6$ at 
    $0.3 < z < 0.5$ where the sample is fairly complete in \mtot{}.
    
    
    
\section{B. Extraction of 1-D surface brightness profile} 
    \label{app:ellipse} 
    
    Here we briefly discuss a few technical issues related to the measurements of the 
    1-D surface brightness profiles around massive galaxies. 
    
\begin{figure*}
        \centering 
        \includegraphics[width=\textwidth]{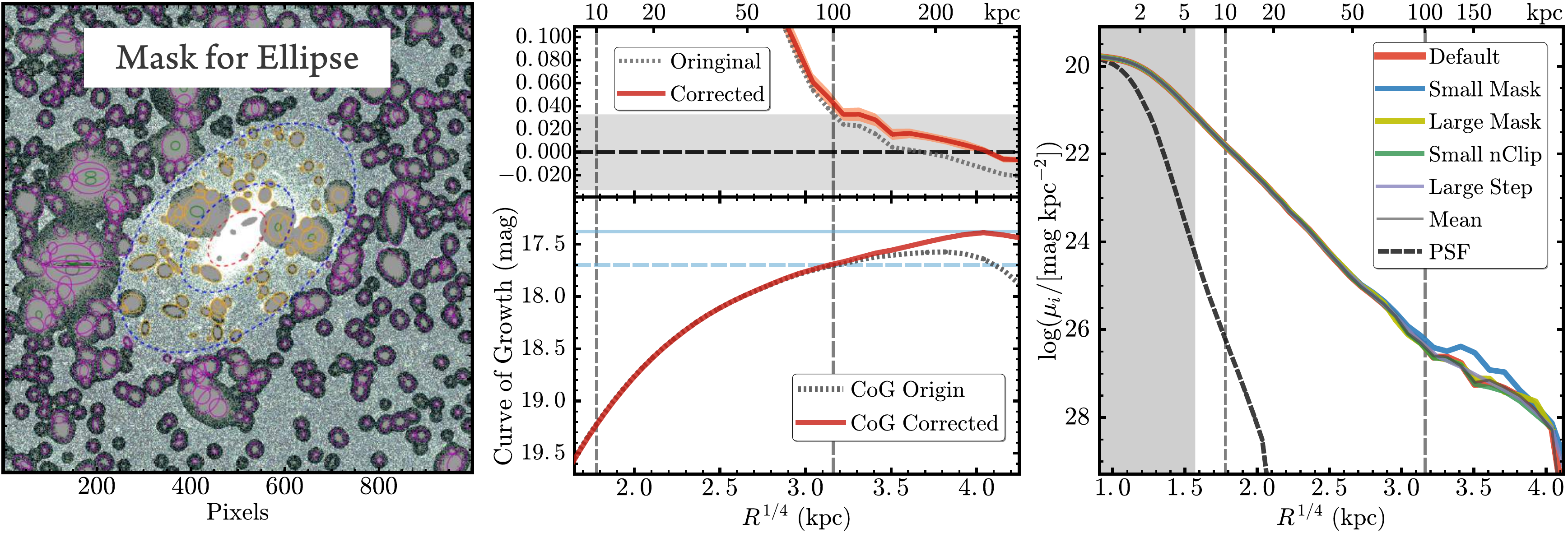}
        \caption{
            \textbf{Left:} Example of the object-mask built for the \texttt{Ellipse}
            run for a typical massive galaxy in the sample. 
            All the shaded regions are masked out. 
            The three dash lines (red, inner one and two blue ones) around the target 
            at the center outlines the three radius we defined using the flux radius 
            of the target.  
            We increase the mask size for objects detected in different regions 
            separated by these apertures (which are outlined by solid, elliptical 
            apertures with different colors) using slightly different criteria.~~
            \textbf{Middle:}: The zoom-in intensity profile around very low intensity 
            value (top panel), and the curve-of-growth of the enclosed magnitude 
            (bottom panel) of the example galaxy.  
            To highlight the importance of background correction, we show the profiles 
            using both images with (red, solid line) and without (black, dotted line) 
            background correction. 
            On the top panel, besides the horizontal line that highlights the zero flux 
            level, we also show the uncertainty of the sky background estimate using 
            the grey-shaded region.  
            On the bottom panel, two horizontal lines indicate the magnitudes 
            corresponding to total flux (solid) and flux within 100 kpc (dash).~~
            \textbf{Right:} compares the 1-D surface brightness profiles for the same 
            example galaxy using different masks 
            (smaller masking region: red, dash line; larger masks: blue, dash line), 
            or different \texttt{Ellipse} configurations
            (more aggressive pixel-clipping: cyan, dash line; 
             larger step in radius: green, dash line; 
             using mean flux along the isophote instead of median: purple, dash line)
            with the default one (black, solid line).
            }
        \label{fig:ell_tech}
    \end{figure*}

    To derive reliable 1-D profile, it is important to mask out all the irrelevant 
    objects around the target.
    At the depth of the HSC images, this becomes a challenging task, especially 
    for massive galaxies with extended outer profiles and many satellites. 
    At this point, the \texttt{hscPipe} tends to over-subtract the background around 
    bright objects.  
    The performance of its deblending process is also not optimized for extended
    objects. 
    For these reasons, we perform \texttt{SExtractor}-like background subtraction and 
    object detection using the \texttt{SEP} Python library to generate the necessary 
    masks.
    Combining two different local background models and $S/N$ thresholds, we obtain 
    the centroid, shape, and radius that enclose 90\% of flux for each object, 
    including the one that is very close to the center of bright galaxy (left panel of 
    Fig~\ref{fig:ell_tech}). 
    Based on these information, we then create the mask that covers all contaminating 
    objects around the target after adaptively increasing the sizes of their masks 
    according to their brightness and distance to the central target. 
    Generally speaking, we mask out bright objects or objects in the outskirt of the 
    image more aggressively to reduce their impact on the surface brightness profiles 
    in the outskirt. 
    We also create masks that are less and more aggressive than the default one to 
    test their impacts on the surface brightness profiles. 
    
    Next, we aggressively mask out all objects on the cut-out image.  
    We then evaluate the background level using the unmasked pixels after median 
    smoothing the masked image using box of $6x6$ pixels.
    This provides estimate of global background level along with its uncertainty. 
    Given the typical background uncertainty, the HSC \texttt{WIDE} image should be 
    able to reach down to $> 29$ \sb surface brightness level in the $i$-band.  
    However, as mentioned, we often find evidence of slightly over-subtracted 
    background for massive galaxies in our sample. 
    In the current \texttt{hscPipe}, the background on each CCD is modeled with a 
    Chebyshev-polynomial that is fit to the smoothed image after excluding pixels 
    with $S/N >5$.
    This algorithm performs much better than the SDSS version 
    (e.g.\ see \citealt{Blanton2011}), yet still over-subtracts background around 
    bright objects and results in unphysical truncation in their surface brightness 
    profiles.
    We empirically correct this issue using the background model generated by 
    the \texttt{SExtractor} algorithm on the masked image 
    ($200x200$ pixels background box size, and 6 pixels median filtering size of 
    sky boxes).
    This model can account for the slightly over-subtracted background at large scale,
    and reduce the impact from the low surface brightness ``wings'' of bright 
    neighbors. 
    We clearly see improvement in both the distributions of background pixels 
    (more symmetric distribution; median value is closer to 0) and the surface 
    brightness profile (middle panel of Fig~\ref{app:ellipse}; the negative intensity 
    and the turn-over of the curve-of-growth in the outskirt of the ``Original'' 
    profile are successfully corrected) after this correction.
    Also, it is worth mentioning that such correction does not often affect the 
    surface brightness profile within 100 kpc. 
    
    The procedure used to derive 1-D surface brightness profile from the 
    background-corrected, contamination-masked images is already described in 
    \S~\ref{sec:ellipse} briefly. 
    In practice, the profile at very low surface brightness level is sensitive to 
    several \texttt{Ellipse} configurations.
    After some tests, we choose to use 0.1 dex in logarithm as the step in semi-major 
    axis length between successive ellipses, and we use the median pixel value over the
    elliptical annulus after rejecting outlying pixels via $3\sigma$-clipping three
    times.
    We make the above choices to make the final profile less affected by any nearby
    object, and also test the differences between the profiles derived using larger
    step, or mean value on the annulus, or fewer times of $\sigma$-clipping. 
    Generally speaking, the surface brightness profile is very robust against these
    changes, especially within 100 kpc. 
    On the right panel of Fig~\ref{fig:ell_tech}, we compare the surface brightness
    profiles for an example massive galaxy using different masks and \texttt{Ellipse}
    parameters. 
    The profile within 100 kpc is very stable, and the only noticable difference 
    is caused by the less aggressive object-mask in the very outskirt.    
    
    We should also mention that we run \texttt{Ellipse} allowing for more 
    sophisticated shapes than simple ellipse (4th Fourier modes that can make 
    isophote more ``disky'' or ``boxy'', e.g.\ \citealt{Kormendy2009}) to fit the 
    isophote better.
    We also apply the isophotes from $i$-band images to other bands in 
    ``force-photometry'' mode \texttt{Ellipse} run to get initial estimates of 
    color profiles.  
    
\section{C. Estimate average {$M_{\star}/L_{\star}$} using \texttt{iSEDFit}} 
    \label{app:sed} 

    \begin{figure*}
        \begin{center}
        \includegraphics[width=\textwidth]{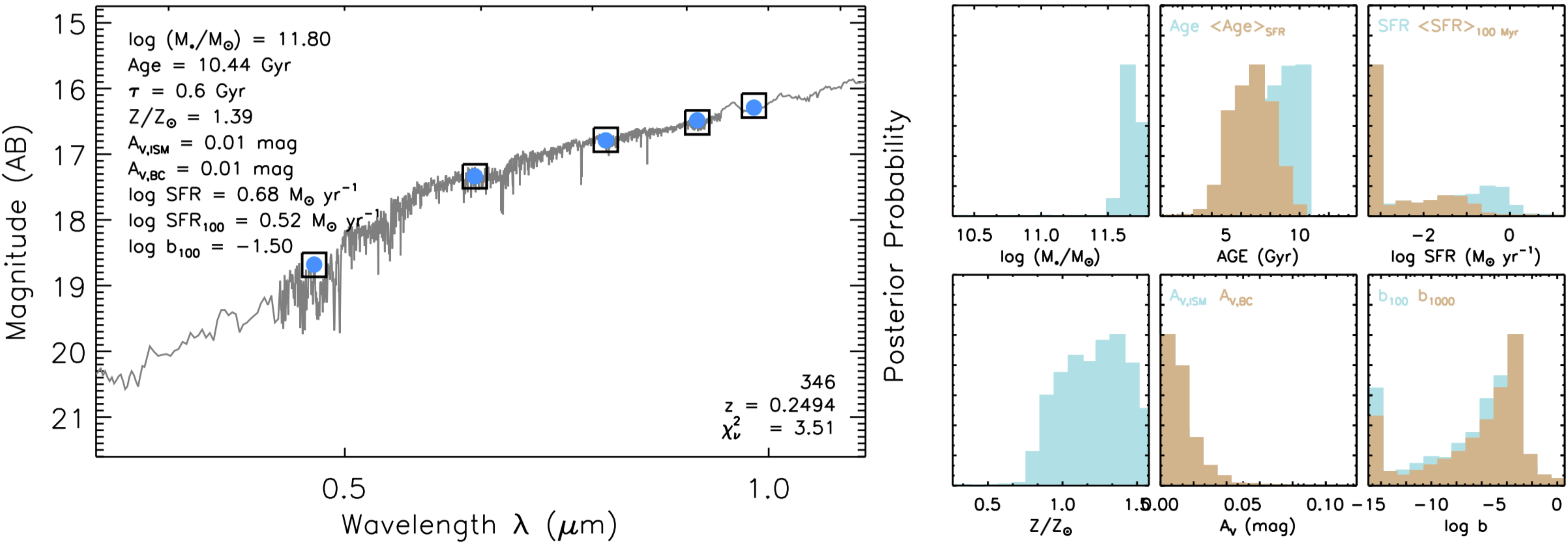}
        \caption{
            \textbf{Left:} Example of output figure from \texttt{iSEDFit} that shows 
            the SED fitting results. 
            The open-boxes show the observed fluxes in 5-band, and the solid, blue-dots
            show the best-fitted results, along with the high-resolution spectrum for
            this model reconstructed using the synthetic spectra from \texttt{FSPS}. 
            Top-left corner shows the best-fit stellar population parameters, and 
            bottom-right corner shows the ID, redshift of this object, and reduced 
            $\chi^2$ of the best-fit model.~~~
            \textbf{Right:} the Posterior distributions of a few key parameters.
            From top-left to bottom right are: 
            1) stellar mass (\logms{}); 
            2) age of the population (mass and star-formation rate weighted) in Gyr; 
            3) star formation rate ($\log\ \mathrm{SFR}\ (M_{\odot}/\mathrm{yr})$; 
            instant one and the one averaged over the previous 100 Myr; 
            4) stellar metallicity ($\mathrm{Z}/\mathrm{Z}_{\odot}$); 
            5) dust extinction ($\mathrm{A}_V$ in mag);
            6) birthrate parameter ($\log\ b$; averaged over previous 100 and 1000 Myr).
            }
        \label{fig:ised}
        \end{center}
    \end{figure*}

    \begin{figure*}
        \begin{center}
        \includegraphics[width=12cm]{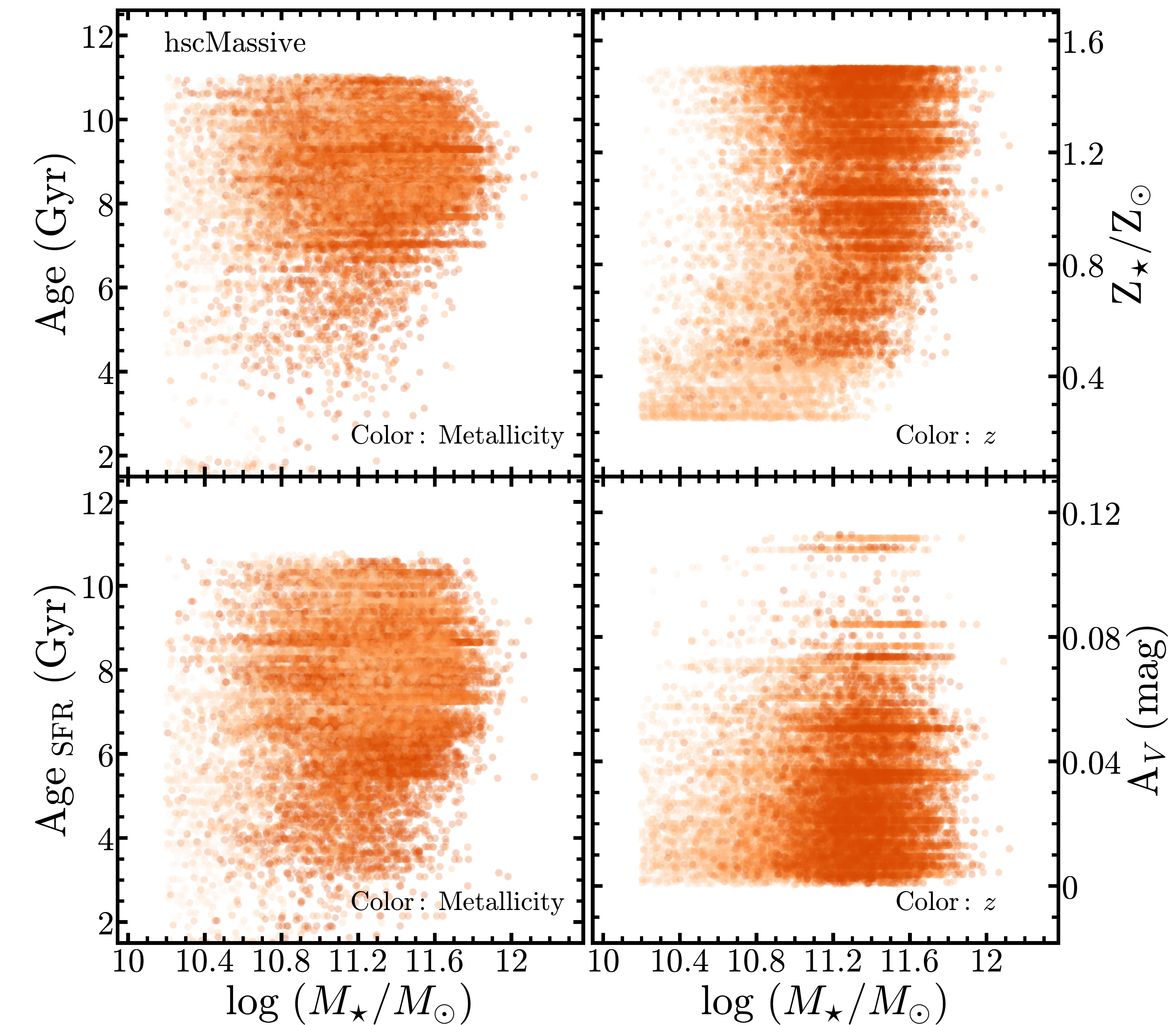}
        \caption{
            Relationships between \mstar{} and key stellar population parameters from 
            \texttt{iSEDfit}. 
            The four stellar population properties are: 
            1) \textbf{Top-left}: \mstar{}-weighted stellar population age in Gyr; 
            2) \textbf{Bottom-left}: SFR-weighted age in Gyr; 
            3) \textbf{Top-right}: \mstar{}-weighted stellar metallicity in unit of 
            Solar value;
            4) \textbf{bottom-right}: dust extinction value in $V$-band.
            As expected, most of the HSC massive galaxies are old, metal-rich, and 
            dust-free. 
            }
        \label{fig:ised_2}
        \end{center}
    \end{figure*}

    \begin{figure}
        \begin{center}
        \includegraphics[width=\columnwidth]{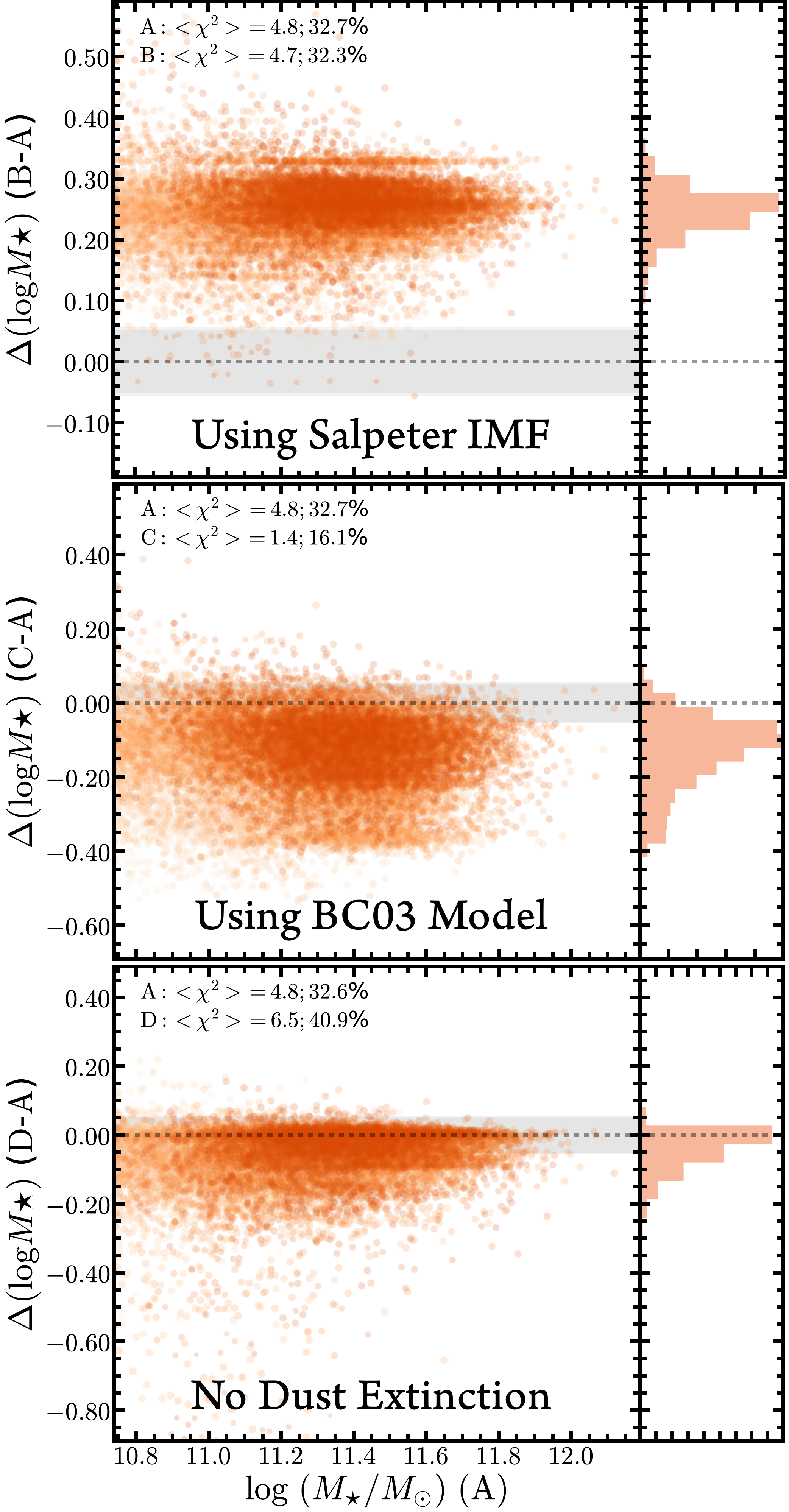}
        \caption{
            Comparisons of \mstar{} estimated by \texttt{iSEDFit} using different
            model assumptions. 
            In each figure, we plot the \mstar{} from the default model against the
            differences with other models. 
            The four models involved are labeled as: 
            (A): Default model; 
            (B): Using the Salpeter IMF instead of the Chabrier one 
                (\textbf{Top panel});
            (C): Using the BC03 synthetic population model instead of the FSPS one
                (\textbf{Middle panel});
            (D): No dust extinction (\textbf{Bottom panel}). 
            On each panel, the grey shaded region highlights the typical uncertainty 
            of the $\log($\logms{}$)$.            
            For each pair of models, we highlight their median $\chi^{2}$ values and 
            the fraction of galaxies with $\chi^{2} > 10.0$ at the top. 
            On each panel, we also show the histograms of the \mstar{}-differences on 
            the right side.
            }
        \label{fig:ised_3}
        \end{center}
    \end{figure}

 
    In \S~\ref{ssec:isedfit}, we briefly explain the SED fitting procedure and 
    the priors used.   
    In Fig~\ref{fig:ised_1}, we show an example of the \texttt{iSEDFit} output by 
    visualizing the 5-band HSC SED on top of the best-fit model along with the PDF 
    of the key parameters.
     
    Although we only use the best-fit \m2l{} in this work, it is necessary to make 
    sure the model is reasonable. 
    We show the relations between \mstar{} and a few key stellar population parameters 
    derived by \texttt{iSEDFit} in Fig~\ref{fig:ised_2}. 
    Degeneracies among these parameters are inevitable based on only five broad-band
    photometry, but as expected, most massive galaxies show old stellar age, high
    stellar metallicity ($1.5 \times Z_{\odot}$ is the highest metallicity allowed by 
    the adopted \texttt{FSPS} SSP models), and low dust extinction.
   
    Meanwhile, \mstar{} measurement based on SED fitting heavily depends on the 
    adopted SSP model, the form of IMF, dust extinction law, and details in 
    the assumption of SFH (e.g.\ \citealt{Bernardi2017}). 
    For massive galaxies in this sample, the form of the SFH\footnote{We choose 
    to use the delayed-$\tau$ model for SFH; we adopt flat distribution between 
    0.5 to 14.0 Gyrs as the prior for the look-back time when the star formation 
    turned on. 
    The exponential delayed time-scale ($\tau$) is allowed to change between 
    0.1 to 3.0 with equal probability}, the contribution from random star 
    burst\footnote{The chance of random star burst is set at 0.2 for every 2 Gyrs. 
    The duration of the star burst is draw from a logarithmic distribution 
    between 0.03 to 0.3 Gyr; and the mass fraction formed in the burst is from 
    a logarithmic distribution between 0.01 and 1.0.} rarely affect the \mstar{}. 
    But he choices of SSP model, IMF, and dust extinction do systematically impact
    the estimates of \mstar{}, and therefore we look into this with a few 
    additional tests (see Fig~\ref{fig:ised_3}):

    \begin{enumerate}

        \item Choosing the \citet{Salpeter1955} IMF results in systematically 
            higher \mstar{} (on average $+0.25$ dex of \logms{}) for these massive 
            galaxies (top panel).
            Although there are multiple lines of evidence that favor Salpeter 
            or even more ``bottom-heavy'' IMF in massive galaxies 
            (e.g.\ \citealt{Conroy2012}; \citealt{Cappellari2012}), we still 
            present the main results using Chabrier IMF to accommodate galaxies 
            with lower \mstar{} in the sample, and to be as consistent as possible 
            with previous works.  
            This choice of IMF does not change the main results qualitatively. 

        \item \mstar{} based on the \texttt{BC03} models are systematically lower 
            than the ones based on \texttt{FSPS+MILES} models (middle panel)
            The difference shows a large scatter, and can be as large as 0.4 dex,
            although it is not \mstar{}-dependent. 
            The \texttt{BC03} results show better average ${\chi}^2$ than the 
            \texttt{FSPS} ones. 
            This relates to the higher upper-limit of stellar metallicity 
            ($2 \times \mathrm{Z}_{\odot}$) allowed by the \texttt{BC03} model, 
            which help fits the shape of the SED in the red-end slightly better.  
            However, the \texttt{BC03} results also show puzzlingly low stellar 
            ages ($< 3$--4 Gyrs) for these massive, red galaxies. 
            This could also lead to underestimated \m2l{} values.
            It is worth noting that, both \texttt{FSPS} and \texttt{BC03} 
            models still have difficulties recovering SED at the very red-end 
            (between $z$ and $y$-band), and reproducing the optical color-color 
            relations for red-sequence galaxies (e.g.\ \citealt{MIUSCAT2}).
            In this work, we decide to keep using the \texttt{FSPS+MILES} model as 
            the fiducial one.  
            Using results based on \texttt{BC03} model will not change any of our 
            conclusions here.
            
        \item On the bottom panel of Fig~\ref{fig:ised_3}, we compare with the
            SED fitting results without considering the dust extinction. 
            This choice leads to slightly smaller \mstar{} values as expected. 
            Its impact becomes slightly larger at lower \mstar{} end. 
            It will not change any of our conclusions here.
          
    \end{enumerate}
     
\clearpage
\include{table1}

\end{document}

%% file: table1.tex
\begin{deluxetable}{c ccc cc cc}
    \label{table:1}
    \centering
    
    \tabletypesize{\scriptsize}
    \tablewidth{0pt}
    \tablecolumns{8}
    \tablenum{1}
    \tablecaption{Average \mden{} Profiles of Massive Galaxies in Different Stellar Mass Bins}
\tablehead{
    \colhead{Radius} & 
    \multicolumn{3}{c}{[\mden{}]; Combined samples} &
    \multicolumn{2}{c}{[\mden{}]; $M_{\star,100\ \mathrm{kpc}}$-matched} &
    \multicolumn{2}{c}{[\mden{}]; $M_{\star,10\ \mathrm{kpc}}$-matched}
	\vspace{1.4ex}
    \nl 
    \colhead{kpc} & 
    \multicolumn{3}{c}{$\log (M_{\odot}/\mathrm{kpc}^2)$} &
    \multicolumn{2}{c}{$\log (M_{\odot}/\mathrm{kpc}^2)$} &
    \multicolumn{2}{c}{$\log (M_{\odot}/\mathrm{kpc}^2)$}
	\vspace{1.4ex}
    \nl 
    \colhead{} & 
    \colhead{$\log \frac{M_{\star,100\mathrm{kpc}}}{M_{\odot}}\in$[11.4, 11.6]} & 
    \colhead{[11.6, 11.8]} & 
    \colhead{[11.8, 12.0]}\hspace{2.0ex} & 
    \colhead{\texttt{cenHighMh}} & 
    \colhead{\texttt{cenLowMh}} & 
    \colhead{\texttt{cenHighMh}}\hspace{2.0ex} & 
    \colhead{\texttt{cenLowMh}}
	\vspace{1.6ex}
    \nl
    \colhead{    (1)} &
    \colhead{    (2)} &
    \colhead{    (3)} &
    \colhead{    (4)} &
    \colhead{    (5)} &
    \colhead{    (6)} &
    \colhead{    (7)} &
    \colhead{    (8)}
}
\startdata

0.0 & $ 9.23\substack{+0.00 \\ -0.00}$ &$ 9.31\substack{+0.00 \\ -0.01}$ &$ 9.32\substack{+0.01 \\ -0.01}$ &$ 9.31\substack{+0.02 \\ -0.02}$ &$ 9.34\substack{+0.01 \\ -0.01}$ &$ 9.31\substack{+0.02 \\ -0.02}$ &$ 9.34\substack{+0.02 \\ -0.02}$ \\
 0.6 & $ 9.20\substack{+0.00 \\ -0.00}$ &$ 9.28\substack{+0.00 \\ -0.01}$ &$ 9.29\substack{+0.01 \\ -0.01}$ &$ 9.27\substack{+0.02 \\ -0.02}$ &$ 9.31\substack{+0.01 \\ -0.01}$ &$ 9.28\substack{+0.02 \\ -0.02}$ &$ 9.31\substack{+0.02 \\ -0.02}$ \\
 1.0 & $ 9.16\substack{+0.00 \\ -0.00}$ &$ 9.24\substack{+0.00 \\ -0.00}$ &$ 9.26\substack{+0.01 \\ -0.01}$ &$ 9.24\substack{+0.02 \\ -0.02}$ &$ 9.27\substack{+0.01 \\ -0.01}$ &$ 9.25\substack{+0.02 \\ -0.02}$ &$ 9.27\substack{+0.02 \\ -0.02}$ \\
 1.4 & $ 9.12\substack{+0.00 \\ -0.00}$ &$ 9.20\substack{+0.00 \\ -0.00}$ &$ 9.23\substack{+0.01 \\ -0.01}$ &$ 9.20\substack{+0.02 \\ -0.02}$ &$ 9.23\substack{+0.01 \\ -0.01}$ &$ 9.21\substack{+0.02 \\ -0.01}$ &$ 9.23\substack{+0.02 \\ -0.01}$ \\
 1.7 & $ 9.06\substack{+0.00 \\ -0.00}$ &$ 9.15\substack{+0.00 \\ -0.00}$ &$ 9.19\substack{+0.01 \\ -0.01}$ &$ 9.15\substack{+0.02 \\ -0.02}$ &$ 9.19\substack{+0.01 \\ -0.01}$ &$ 9.16\substack{+0.01 \\ -0.01}$ &$ 9.18\substack{+0.01 \\ -0.01}$ \\
 2.0 & $ 9.00\substack{+0.00 \\ -0.00}$ &$ 9.10\substack{+0.00 \\ -0.00}$ &$ 9.15\substack{+0.01 \\ -0.01}$ &$ 9.09\substack{+0.01 \\ -0.02}$ &$ 9.13\substack{+0.01 \\ -0.01}$ &$ 9.11\substack{+0.01 \\ -0.01}$ &$ 9.12\substack{+0.01 \\ -0.01}$ \\
 2.4 & $ 8.93\substack{+0.00 \\ -0.00}$ &$ 9.03\substack{+0.00 \\ -0.00}$ &$ 9.09\substack{+0.01 \\ -0.01}$ &$ 9.03\substack{+0.02 \\ -0.02}$ &$ 9.07\substack{+0.01 \\ -0.01}$ &$ 9.05\substack{+0.01 \\ -0.01}$ &$ 9.05\substack{+0.01 \\ -0.01}$ \\
 2.7 & $ 8.87\substack{+0.00 \\ -0.00}$ &$ 8.97\substack{+0.00 \\ -0.00}$ &$ 9.04\substack{+0.01 \\ -0.01}$ &$ 8.97\substack{+0.01 \\ -0.01}$ &$ 9.01\substack{+0.01 \\ -0.01}$ &$ 9.00\substack{+0.01 \\ -0.01}$ &$ 8.99\substack{+0.01 \\ -0.01}$ \\
 3.0 & $ 8.80\substack{+0.00 \\ -0.00}$ &$ 8.90\substack{+0.00 \\ -0.00}$ &$ 8.98\substack{+0.01 \\ -0.01}$ &$ 8.90\substack{+0.01 \\ -0.01}$ &$ 8.95\substack{+0.01 \\ -0.01}$ &$ 8.93\substack{+0.01 \\ -0.01}$ &$ 8.92\substack{+0.01 \\ -0.01}$ \\
 3.4 & $ 8.72\substack{+0.00 \\ -0.00}$ &$ 8.83\substack{+0.00 \\ -0.00}$ &$ 8.92\substack{+0.01 \\ -0.01}$ &$ 8.83\substack{+0.01 \\ -0.01}$ &$ 8.88\substack{+0.01 \\ -0.01}$ &$ 8.86\substack{+0.01 \\ -0.01}$ &$ 8.85\substack{+0.01 \\ -0.01}$ \\
 3.7 & $ 8.66\substack{+0.00 \\ -0.00}$ &$ 8.78\substack{+0.00 \\ -0.00}$ &$ 8.87\substack{+0.01 \\ -0.01}$ &$ 8.78\substack{+0.01 \\ -0.01}$ &$ 8.83\substack{+0.01 \\ -0.01}$ &$ 8.81\substack{+0.01 \\ -0.01}$ &$ 8.79\substack{+0.01 \\ -0.01}$ \\
 4.1 & $ 8.60\substack{+0.00 \\ -0.00}$ &$ 8.72\substack{+0.00 \\ -0.00}$ &$ 8.82\substack{+0.01 \\ -0.01}$ &$ 8.72\substack{+0.01 \\ -0.01}$ &$ 8.77\substack{+0.01 \\ -0.01}$ &$ 8.76\substack{+0.01 \\ -0.01}$ &$ 8.73\substack{+0.01 \\ -0.01}$ \\
 4.4 & $ 8.54\substack{+0.00 \\ -0.00}$ &$ 8.66\substack{+0.00 \\ -0.00}$ &$ 8.77\substack{+0.01 \\ -0.01}$ &$ 8.66\substack{+0.01 \\ -0.01}$ &$ 8.72\substack{+0.01 \\ -0.01}$ &$ 8.70\substack{+0.01 \\ -0.01}$ &$ 8.67\substack{+0.01 \\ -0.01}$ \\
 4.8 & $ 8.48\substack{+0.00 \\ -0.00}$ &$ 8.60\substack{+0.00 \\ -0.00}$ &$ 8.71\substack{+0.01 \\ -0.01}$ &$ 8.60\substack{+0.01 \\ -0.01}$ &$ 8.66\substack{+0.01 \\ -0.01}$ &$ 8.65\substack{+0.01 \\ -0.01}$ &$ 8.61\substack{+0.01 \\ -0.01}$ \\
 6.2 & $ 8.26\substack{+0.00 \\ -0.00}$ &$ 8.40\substack{+0.00 \\ -0.00}$ &$ 8.53\substack{+0.01 \\ -0.01}$ &$ 8.41\substack{+0.01 \\ -0.01}$ &$ 8.46\substack{+0.01 \\ -0.01}$ &$ 8.46\substack{+0.02 \\ -0.02}$ &$ 8.40\substack{+0.02 \\ -0.02}$ \\
 7.6 & $ 8.09\substack{+0.00 \\ -0.00}$ &$ 8.24\substack{+0.00 \\ -0.00}$ &$ 8.39\substack{+0.01 \\ -0.01}$ &$ 8.27\substack{+0.01 \\ -0.01}$ &$ 8.31\substack{+0.01 \\ -0.01}$ &$ 8.31\substack{+0.02 \\ -0.02}$ &$ 8.23\substack{+0.02 \\ -0.02}$ \\
 9.0 & $ 7.95\substack{+0.00 \\ -0.00}$ &$ 8.10\substack{+0.00 \\ -0.00}$ &$ 8.27\substack{+0.01 \\ -0.01}$ &$ 8.14\substack{+0.02 \\ -0.02}$ &$ 8.18\substack{+0.01 \\ -0.01}$ &$ 8.19\substack{+0.02 \\ -0.02}$ &$ 8.09\substack{+0.02 \\ -0.02}$ \\
10.3 & $ 7.82\substack{+0.00 \\ -0.00}$ &$ 7.99\substack{+0.00 \\ -0.00}$ &$ 8.16\substack{+0.01 \\ -0.01}$ &$ 8.03\substack{+0.02 \\ -0.01}$ &$ 8.06\substack{+0.01 \\ -0.01}$ &$ 8.09\substack{+0.02 \\ -0.02}$ &$ 7.97\substack{+0.02 \\ -0.02}$ \\
11.7 & $ 7.70\substack{+0.00 \\ -0.00}$ &$ 7.88\substack{+0.00 \\ -0.00}$ &$ 8.06\substack{+0.01 \\ -0.01}$ &$ 7.93\substack{+0.02 \\ -0.02}$ &$ 7.96\substack{+0.01 \\ -0.01}$ &$ 7.99\substack{+0.02 \\ -0.02}$ &$ 7.85\substack{+0.02 \\ -0.02}$ \\
13.0 & $ 7.60\substack{+0.00 \\ -0.00}$ &$ 7.78\substack{+0.00 \\ -0.00}$ &$ 7.98\substack{+0.01 \\ -0.01}$ &$ 7.85\substack{+0.02 \\ -0.02}$ &$ 7.87\substack{+0.01 \\ -0.01}$ &$ 7.90\substack{+0.02 \\ -0.02}$ &$ 7.75\substack{+0.02 \\ -0.02}$ \\
14.5 & $ 7.50\substack{+0.00 \\ -0.00}$ &$ 7.69\substack{+0.00 \\ -0.00}$ &$ 7.90\substack{+0.01 \\ -0.01}$ &$ 7.76\substack{+0.02 \\ -0.02}$ &$ 7.78\substack{+0.01 \\ -0.01}$ &$ 7.82\substack{+0.02 \\ -0.02}$ &$ 7.65\substack{+0.02 \\ -0.02}$ \\
16.0 & $ 7.39\substack{+0.00 \\ -0.00}$ &$ 7.60\substack{+0.00 \\ -0.00}$ &$ 7.82\substack{+0.01 \\ -0.01}$ &$ 7.68\substack{+0.02 \\ -0.02}$ &$ 7.69\substack{+0.01 \\ -0.01}$ &$ 7.74\substack{+0.02 \\ -0.03}$ &$ 7.56\substack{+0.02 \\ -0.03}$ \\
17.3 & $ 7.31\substack{+0.00 \\ -0.00}$ &$ 7.52\substack{+0.00 \\ -0.00}$ &$ 7.76\substack{+0.01 \\ -0.01}$ &$ 7.61\substack{+0.02 \\ -0.02}$ &$ 7.62\substack{+0.01 \\ -0.01}$ &$ 7.67\substack{+0.03 \\ -0.03}$ &$ 7.48\substack{+0.03 \\ -0.03}$ \\
18.7 & $ 7.23\substack{+0.00 \\ -0.00}$ &$ 7.45\substack{+0.00 \\ -0.00}$ &$ 7.69\substack{+0.01 \\ -0.01}$ &$ 7.55\substack{+0.02 \\ -0.02}$ &$ 7.55\substack{+0.01 \\ -0.01}$ &$ 7.61\substack{+0.03 \\ -0.03}$ &$ 7.40\substack{+0.03 \\ -0.03}$ \\
22.6 & $ 7.02\substack{+0.00 \\ -0.00}$ &$ 7.27\substack{+0.00 \\ -0.00}$ &$ 7.54\substack{+0.01 \\ -0.01}$ &$ 7.38\substack{+0.02 \\ -0.02}$ &$ 7.37\substack{+0.01 \\ -0.01}$ &$ 7.45\substack{+0.03 \\ -0.03}$ &$ 7.21\substack{+0.03 \\ -0.03}$ \\
26.1 & $ 6.86\substack{+0.00 \\ -0.00}$ &$ 7.12\substack{+0.00 \\ -0.00}$ &$ 7.41\substack{+0.01 \\ -0.01}$ &$ 7.25\substack{+0.02 \\ -0.02}$ &$ 7.24\substack{+0.01 \\ -0.01}$ &$ 7.32\substack{+0.03 \\ -0.03}$ &$ 7.05\substack{+0.03 \\ -0.03}$ \\
30.0 & $ 6.70\substack{+0.00 \\ -0.00}$ &$ 6.98\substack{+0.00 \\ -0.00}$ &$ 7.29\substack{+0.01 \\ -0.01}$ &$ 7.13\substack{+0.03 \\ -0.02}$ &$ 7.10\substack{+0.01 \\ -0.01}$ &$ 7.20\substack{+0.03 \\ -0.04}$ &$ 6.90\substack{+0.03 \\ -0.04}$ \\
33.7 & $ 6.55\substack{+0.00 \\ -0.00}$ &$ 6.85\substack{+0.01 \\ -0.01}$ &$ 7.18\substack{+0.01 \\ -0.01}$ &$ 7.01\substack{+0.03 \\ -0.03}$ &$ 6.98\substack{+0.01 \\ -0.01}$ &$ 7.09\substack{+0.03 \\ -0.03}$ &$ 6.76\substack{+0.03 \\ -0.03}$ \\
37.8 & $ 6.41\substack{+0.00 \\ -0.00}$ &$ 6.72\substack{+0.01 \\ -0.01}$ &$ 7.07\substack{+0.01 \\ -0.01}$ &$ 6.90\substack{+0.03 \\ -0.03}$ &$ 6.85\substack{+0.01 \\ -0.01}$ &$ 6.98\substack{+0.04 \\ -0.04}$ &$ 6.63\substack{+0.04 \\ -0.04}$ \\
41.6 & $ 6.29\substack{+0.01 \\ -0.01}$ &$ 6.61\substack{+0.01 \\ -0.01}$ &$ 6.98\substack{+0.01 \\ -0.01}$ &$ 6.81\substack{+0.03 \\ -0.03}$ &$ 6.75\substack{+0.01 \\ -0.01}$ &$ 6.89\substack{+0.04 \\ -0.04}$ &$ 6.51\substack{+0.04 \\ -0.04}$ \\
45.7 & $ 6.17\substack{+0.01 \\ -0.01}$ &$ 6.50\substack{+0.01 \\ -0.01}$ &$ 6.88\substack{+0.01 \\ -0.01}$ &$ 6.71\substack{+0.03 \\ -0.03}$ &$ 6.64\substack{+0.01 \\ -0.01}$ &$ 6.79\substack{+0.04 \\ -0.04}$ &$ 6.39\substack{+0.04 \\ -0.04}$ \\
49.3 & $ 6.07\substack{+0.01 \\ -0.01}$ &$ 6.41\substack{+0.01 \\ -0.01}$ &$ 6.80\substack{+0.01 \\ -0.02}$ &$ 6.62\substack{+0.03 \\ -0.03}$ &$ 6.56\substack{+0.01 \\ -0.01}$ &$ 6.70\substack{+0.04 \\ -0.04}$ &$ 6.30\substack{+0.04 \\ -0.04}$ \\
53.1 & $ 5.98\substack{+0.01 \\ -0.01}$ &$ 6.33\substack{+0.01 \\ -0.01}$ &$ 6.71\substack{+0.02 \\ -0.02}$ &$ 6.55\substack{+0.03 \\ -0.03}$ &$ 6.46\substack{+0.01 \\ -0.01}$ &$ 6.64\substack{+0.04 \\ -0.04}$ &$ 6.21\substack{+0.04 \\ -0.04}$ \\
57.2 & $ 5.88\substack{+0.01 \\ -0.01}$ &$ 6.24\substack{+0.01 \\ -0.01}$ &$ 6.63\substack{+0.02 \\ -0.02}$ &$ 6.47\substack{+0.04 \\ -0.04}$ &$ 6.37\substack{+0.01 \\ -0.01}$ &$ 6.56\substack{+0.04 \\ -0.04}$ &$ 6.11\substack{+0.04 \\ -0.04}$ \\
61.5 & $ 5.79\substack{+0.01 \\ -0.01}$ &$ 6.15\substack{+0.01 \\ -0.01}$ &$ 6.55\substack{+0.02 \\ -0.02}$ &$ 6.39\substack{+0.04 \\ -0.04}$ &$ 6.29\substack{+0.01 \\ -0.01}$ &$ 6.49\substack{+0.04 \\ -0.04}$ &$ 6.03\substack{+0.04 \\ -0.04}$ \\
66.0 & $ 5.70\substack{+0.01 \\ -0.01}$ &$ 6.05\substack{+0.01 \\ -0.01}$ &$ 6.47\substack{+0.02 \\ -0.02}$ &$ 6.32\substack{+0.04 \\ -0.04}$ &$ 6.20\substack{+0.01 \\ -0.01}$ &$ 6.37\substack{+0.05 \\ -0.06}$ &$ 5.94\substack{+0.05 \\ -0.06}$ \\
69.8 & $ 5.64\substack{+0.01 \\ -0.01}$ &$ 5.98\substack{+0.01 \\ -0.01}$ &$ 6.40\substack{+0.02 \\ -0.02}$ &$ 6.25\substack{+0.04 \\ -0.04}$ &$ 6.12\substack{+0.02 \\ -0.01}$ &$ 6.35\substack{+0.04 \\ -0.05}$ &$ 5.87\substack{+0.04 \\ -0.05}$ \\
74.7 & $ 5.56\substack{+0.01 \\ -0.01}$ &$ 5.89\substack{+0.01 \\ -0.01}$ &$ 6.32\substack{+0.02 \\ -0.02}$ &$ 6.18\substack{+0.04 \\ -0.04}$ &$ 6.04\substack{+0.02 \\ -0.02}$ &$ 6.28\substack{+0.05 \\ -0.05}$ &$ 5.79\substack{+0.05 \\ -0.05}$ \\
79.9 & $ 5.49\substack{+0.01 \\ -0.01}$ &$ 5.81\substack{+0.01 \\ -0.01}$ &$ 6.24\substack{+0.02 \\ -0.02}$ &$ 6.12\substack{+0.04 \\ -0.04}$ &$ 5.96\substack{+0.02 \\ -0.02}$ &$ 6.20\substack{+0.05 \\ -0.06}$ &$ 5.72\substack{+0.05 \\ -0.06}$ \\
84.3 & $ 5.43\substack{+0.01 \\ -0.01}$ &$ 5.74\substack{+0.01 \\ -0.01}$ &$ 6.18\substack{+0.02 \\ -0.02}$ &$ 6.05\substack{+0.04 \\ -0.05}$ &$ 5.89\substack{+0.02 \\ -0.02}$ &$ 6.16\substack{+0.05 \\ -0.05}$ &$ 5.65\substack{+0.05 \\ -0.05}$ \\
88.8 & $ 5.38\substack{+0.01 \\ -0.01}$ &$ 5.67\substack{+0.01 \\ -0.01}$ &$ 6.11\substack{+0.02 \\ -0.02}$ &$ 5.99\substack{+0.05 \\ -0.06}$ &$ 5.81\substack{+0.02 \\ -0.02}$ &$ 6.08\substack{+0.05 \\ -0.06}$ &$ 5.58\substack{+0.05 \\ -0.06}$ \\
97.2 & $ 5.29\substack{+0.01 \\ -0.01}$ &$ 5.56\substack{+0.01 \\ -0.01}$ &$ 5.98\substack{+0.02 \\ -0.02}$ &$ 5.92\substack{+0.04 \\ -0.04}$ &$ 5.69\substack{+0.02 \\ -0.02}$ &$ 5.99\substack{+0.05 \\ -0.05}$ &$ 5.47\substack{+0.05 \\ -0.05}$ \\
103.6 & $ 5.21\substack{+0.01 \\ -0.01}$ &$ 5.49\substack{+0.01 \\ -0.01}$ &$ 5.89\substack{+0.03 \\ -0.03}$ &$ 5.84\substack{+0.05 \\ -0.05}$ &$ 5.62\substack{+0.02 \\ -0.02}$ &$ 5.94\substack{+0.05 \\ -0.05}$ &$ 5.39\substack{+0.05 \\ -0.05}$ \\
111.6 & $ 5.14\substack{+0.01 \\ -0.01}$ &$ 5.40\substack{+0.01 \\ -0.01}$ &$ 5.79\substack{+0.03 \\ -0.03}$ &$ 5.78\substack{+0.05 \\ -0.05}$ &$ 5.54\substack{+0.02 \\ -0.02}$ &$ 5.87\substack{+0.05 \\ -0.05}$ &$ 5.32\substack{+0.05 \\ -0.05}$ \\
117.2 & $ 5.10\substack{+0.01 \\ -0.01}$ &$ 5.36\substack{+0.01 \\ -0.01}$ &$ 5.72\substack{+0.03 \\ -0.03}$ &$ 5.72\substack{+0.05 \\ -0.05}$ &$ 5.47\substack{+0.02 \\ -0.02}$ &$ 5.82\substack{+0.05 \\ -0.05}$ &$ 5.29\substack{+0.05 \\ -0.05}$ \\
129.0 & $ 5.00\substack{+0.01 \\ -0.01}$ &$ 5.25\substack{+0.02 \\ -0.02}$ &$ 5.61\substack{+0.03 \\ -0.03}$ &$ 5.64\substack{+0.05 \\ -0.05}$ &$ 5.36\substack{+0.02 \\ -0.02}$ &$ 5.74\substack{+0.05 \\ -0.05}$ &$ 5.21\substack{+0.05 \\ -0.05}$ \\
141.7 & $ 4.89\substack{+0.02 \\ -0.02}$ &$ 5.13\substack{+0.02 \\ -0.02}$ &$ 5.49\substack{+0.03 \\ -0.03}$ &$ 5.58\substack{+0.05 \\ -0.05}$ &$ 5.23\substack{+0.03 \\ -0.03}$ &$ 5.66\substack{+0.05 \\ -0.05}$ &$ 5.09\substack{+0.05 \\ -0.05}$ \\
146.7 & $ 4.85\substack{+0.02 \\ -0.02}$ &$ 5.10\substack{+0.02 \\ -0.02}$ &$ 5.46\substack{+0.03 \\ -0.03}$ &$ 5.51\substack{+0.06 \\ -0.06}$ &$ 5.19\substack{+0.03 \\ -0.03}$ &$ 5.61\substack{+0.05 \\ -0.05}$ &$ 5.03\substack{+0.05 \\ -0.05}$ \\

\enddata
\tablecomments{
    Average \mden{} profiles of massive \rbcg{} and \nbcg{} galaxies in different
    samples:\\ 
    Col.~(1) Radius along the major axis in kpc.\\
    Col.~(2) Average \mden{} profile for galaxies with 
        $11.4 \leq$\logmtot$< 11.6$ in the combined samples of \rbcg{} and \nbcg{}
        galaxies. \\ 
    Col.~(3) Average \mden{} profile of combined samples in the mass bin of 
        $11.6 \leq$\logmtot$< 11.8$. \\ 
    Col.~(4) Average \mden{} profile of combined samples in the mass bin of 
        $11.8 \leq$\logmtot$< 12.0$. \\ 
    Col.~(5) and Col.~(6) are the average \mden{} profiles of \rbcg{} and \nbcg{} galaxies
        in the \mtot{}-matched samples within $11.6 \leq$\logmtot{}$< 11.9$. \\ 
    Col.~(7) and Col.~(8) are the average \mden{} profiles of \rbcg{} and \nbcg{} galaxies 
        in the \minn{}-matched samples within $11.2 \leq$\logmtot{}$< 11.6$. \\ 
    The upper and lower uncertainties of these average profiles vial bootstrap-resampling 
    method are also displayed.
}
\label{tab:prof}
\end{deluxetable}